\documentclass[prd,superscriptaddress,altaffilletter,showpacs,nofootinbib]{revtex4}
\usepackage[dvips]{graphicx}
\usepackage{amsmath}
\newcommand{\be}{\begin{equation}}
\newcommand{\ee}{\end{equation}}
\newcommand{\bea}{\begin{eqnarray}}
\newcommand{\eea}{\end{eqnarray}}
\newcommand{\der}{\partial}
\newcommand{\vphi}{\varphi}

\begin{document}

\title{Dynamical systems analysis of the cubic galileon beyond the exponential potential and the cosmological analogue of the vDVZ discontinuity}

\author{Roberto De Arcia}\email{robertodearcia@gmail.com}\affiliation{Dpto. Ingenier\'ia Civil, Divisi\'on de Ingenier\'ia, Universidad de Guanajuato, Gto., M\'exico.}

\author{Tame Gonzalez}\email{tamegc72@gmail.com}\affiliation{Dpto. Ingenier\'ia Civil, Divisi\'on de Ingenier\'ia, Universidad de Guanajuato, Gto., M\'exico.}

\author{Francisco Antonio Horta-Rangel}\email{anthort@hotmail.com}\affiliation{Dpto. Ingenier\'ia Civil, Divisi\'on de Ingenier\'ia, Universidad de Guanajuato, Gto., M\'exico.}

\author{Genly Leon}\email{genly.leon@ucn.cl}\affiliation{Departamento de Matem\'aticas, Universidad Cat\'olica del Norte, Avda. Angamos 0610, Casilla 1280 Antofagasta, Chile.}

\author{Ulises Nucamendi}\email{ulises@ifm.umich.mx}\affiliation{Instituto de F\'isica y Matem\'aticas, Universidad Michoacana de San Nicol\'as de Hidalgo, Edificio C-3, Ciudad Universitaria, CP. 58040 Morelia, Michoac\'an, M\'exico.}

\author{Israel Quiros}\email{iquiros@fisica.ugto.mx}\affiliation{Dpto. Ingenier\'ia Civil, Divisi\'on de Ingenier\'ia, Universidad de Guanajuato, Gto., M\'exico.}

\begin{abstract}
In this paper we generalize the dynamical systems analysis of the cubic galileon model previously investigated in \cite{rtgui} by including self-interaction potentials beyond the exponential one. It will be shown that, consistently with the results of \cite{rtgui}, the cubic self-interaction of the galileon vacuum appreciably modifies the late-time cosmic dynamics by the existence of a phantom-like attractor (among other super-accelerated solutions that do not modify in any appreciable way the late-time dynamics and hence are not of interest in the present investigation). In contrast, in the presence of background matter the late-time cosmic dynamics remains practically the same as in the standard quintessence scenario. This means that we can not recover the cubic galileon vacuum continuously from the more general cubic quintessence with background matter, by setting to zero the matter energy density (and the pressure). This happens to be a kind of cosmological vDVZ discontinuity that can be evaded by means of the cosmological version of the Vainshtein screening mechanism.
\end{abstract}


\maketitle

%\date{\today}

\section{Introduction}\label{sec-intro}

According to the increasing set of independent cosmological observations \cite{Riess, snia, bao, cmb, lss, lensing, swe} the universe today is experiencing an accelerated expansion era. An unknown component dubbed as dark energy has been proposed to explain this recent acceleration in the context of the general relativity. The cosmological constant with equation of state $\omega = -1$, is the simplest and the most accurate candidate according the observations \cite{lambda}. However, it is plagued  by serious theoretical issues such as the vacuum energy problem, the cosmic coincidence problem, the particle nature of dark matter, the validity of general relativity on large scales, and the age problem \cite{peebles, age}. Since the observations allow the variation in time of the dark energy component, another possibility is to consider the existence of  light scalar fields known as ``quintessence''\cite{ratra}. 

Modified gravity represents an alternative approach for addressing the unusual cosmological dynamics at large scales. It is based on the modification of general relativity. We can observe two main streams in this context: introducing a Lagrangian built up of a Ricci, Riemann or another metric tensors as in the case of $f(R, \mathcal{G})$ theories \cite{Capozziello} and Brans-Dicke (BD) theories \cite{Brans}, or assuming the existence of additional dimensions that realize cosmic acceleration through the leakage of gravity into the extra-space at cosmological scales as in the  Dvali-Gabadadze-Porrati (DGP) braneworld \cite{dgp, roy-rev}. This latter model, however, is plagued by ghost instabilities that cast doubts on its validity.\footnote{As a matter of fact the ghost instability in the DGP model arise only for the self-accelerating branch of the solution, i.e., the one of physical interest to expain the late-time acceleration of the cosmic expansion.} 

Inspired by the DGP model, in \cite{nicolis} the authors proposed an infrared modification of gravity which is a generalization of the 4D effective theory in the DGP braneworld. The theory is invariant under the Galilean shift symmetry $\der_\mu\phi\rightarrow\der_\mu\phi+b_\mu$ in the Minkowski space-time, which keeps the equations of motion at second order. The scalar field that respects the Galilean symmetry is dubbed ``galileon''. The model has a self-accelerating de Sitter solution with no ghost-like instability. The analysis in \cite{nicolis} is valid only for weak gravity in flat spacetime, so that the above result must change in the covariant version of the model \cite{deffayet, deffayet-rev, chow}. As a matter of fact, in the covariantized theory of the galileon the shift symmetry is not preserved, however, the equations of motion still are second order, which is primordial since the higher-derivative theories are in general plagued by the so called Ostrogradsky instability \cite{ostro-inst}. 

In \cite{silva-kazuya} a covariant Brans-Dicke galileon model exhibiting the self-accelerating solution was proposed that was free of ghostlike instabilities. The key feature in the model was the cubic self-interaction term of the form $f(\phi)\nabla^2\phi(\der\phi)^2$. This is the unique form of interactions at cubic order yielding a second-order motion equation for the galileon field. A related Einstein's frame (EF) cubic galileon model given by the Lagrangian\footnote{A similar Lagrangian is found in \cite{chow}.} $${\cal L}=-3(\der\phi)^2-\frac{1}{\Lambda^3}\nabla^2\phi(\der\phi)^2+\frac{g}{M_\text{Pl}}\phi\,T,$$ where $(\der\phi)^2\equiv g^{\mu\nu}\der_\mu\phi\der_\nu\phi$, $\nabla^2\phi\equiv g^{\mu\nu}\nabla_\mu\nabla_\nu\phi$, $g\sim{\cal O}(1)$ for gravitational strength coupling, $M_\text{Pl}$ is the Planck scale, $\Lambda$ is the strong-coupling of the theory and $T\equiv g^{\mu\nu}T_{\mu\nu}$ is the trace of the stress-energy tensor of matter, provides the simplest non-trivial theory exhibiting the Vainshtein screening mechanism \cite{vainshtein, khoury-rev}. The Vainshtein mechanism relies on the cubic self-interaction term $\nabla^2\phi(\der\phi)^2/\Lambda^3$ becoming large compared to the kinetic term $(\der\phi)^2$ near massive objects. The above cubic galileon Lagrangian belongs in the wider class of the so called Horndeski theories \cite{horndeski} that represent the generalization of scalar-tensor theories to include higher-derivative terms. The motion equations for Horndeski theories are second order, thus warranting the absence of the Ostrogradsky instability.

A simplified cubic galileon model of cosmological interest is given by the following EF action:

\bea S=\int\frac{d^4x\sqrt{-g}}{2}\left\{R-\left[1+\sigma(\nabla^2\phi)\right](\nabla\phi)^2-2V\right\}+\int d^4x\sqrt{-g}\,{\cal L}_m,\label{action}\eea where $V=V(\phi)$ is the self-interacting potential of the scalar field, $\sigma=\sigma(\phi)$ is a coupling function and ${\cal L}_m$ stands for the matter Lagrangian (here we have chosen the units where $8\pi G_N=c=1$). This scenario has been probed to be very interesting and has been studied in detail in the literature. In \cite{genly} a dynamical systems study of the model \eqref{action} with the inclusion of the background matter (${\cal L}_m$) was developed for a pair of self-interaction potentials, showing that the cubic self-interaction of the galileon has no impact in the late-time cosmic dynamics. 

A very interesting result was obtained in \cite{rtgui} for the cubic galileon \eqref{action} with the exponential self-interaction potential: When the matter degrees of freedom other than the galileon itself are removed -- i. e., when the vacuum galileon action \eqref{action} with ${\cal L}_m=0$ is considered -- the late time dynamics can be indeed modified by the presence of a phantom attractor associated with super-accelerated expansion, a result that has no analogue in the case when the matter Lagrangian ${\cal L}_m$ is considered. Hence it results that one can not recover the cubic galileon vacuum dynamics continuously from the more general case with the inclusion of matter by setting to zero the matter energy density (and the pressure). We argue that this is a kind of cosmological vDVZ discontinuity \cite{vdvz} that, as in similar cases in the bibliography, can be evaded by means of the cosmological analogue of the Vainshtein screening mechanism \cite{chow, silva-kazuya, vainshtein, khoury-rev} that is due to the cubic term in \eqref{action}: $\sigma\nabla^2\phi(\der\phi)^2$. It happens that, as the effective (phantom-like) energy density grows up with the cosmic expansion, the cubic self-interaction term dominates the dynamics of the expansion, leading to the decoupling of the galileon from the remaining degrees of freedom, and to the eventual recovering of general relativity. This is how, in the presence of background matter, the cosmological Vainshtein screening mechanism prevents the occurrence of a big-rip singularity a finite time into the future in the present cubic galileon model. In other words, in the presence of standard matter, the phantom attractor arising in the vacuum case is erased from the phase space by means of the Vainshtein-like screening, a fact that is consistent with the result of \cite{genly} that the late-time dynamics of the model \eqref{action} with the presence of background matter is basically the same as for the standard quintessence.\footnote{A related interesting model is the one studied in the reference \cite{Dimakis:2017kwx} that can be seen as an special case of the model \cite{genly}, but not covered there. In that reference a class of generalized Galileon cosmological models, which can be described by a point-like Lagrangian, is considered in order to utilize Noether's theorem to determine conservation laws for the field equations. In the Friedmann-Robertson-Walker (FRW) universe, the existence of a nontrivial conservation law indicates the integrability of equations. Due to the complexity of the latter, the authors apply the differential invariants approach in order to construct special power-law solutions and study their stability. Approximated solutions, similar to these exact solutions with power-law behaviors $a=a_0 t^{p}$ are also found in the present paper as we will see later on.}

The above result was obtained for a particular choice of the self-interaction potential: the exponential potential. In the present paper we shall go further to show that the above kind of cosmological vDVZ discontinuity -- and its resolution through the cosmological version of the Vainshtein screening effect -- is independent of the specific form of the potential $V$. Although we do not aim at demonstrating this for the general case, but for several choices of the self-interaction potential beyond the exponential one (see below), our results are quite general since several of the critical points found exist for arbitrary potentials. 

The plan of the paper is the following: In section \ref{sec2} we introduce the cosmological equations that govern the dynamics of the model and we also compute helpful cosmological parameters. The self-interaction potentials beyond the exponential one that will be investigated in this paper are presented in section \ref{sec-bey-exp}. Then in section \ref{sec3} we expose the generalities of the dynamical system corresponding to the cosmological model of interest in the equivalent phase space. The phase space dynamics of the generalized galileon model with cubic derivative interaction in the presence of background matter is analyzed in section \ref{sec4}. In section \ref{sec5} we focus in an apparently simpler case: the pure galileon vacuum. We are able to confirm that the cosmological dynamics of the vacuum is richer than the one in the presence of background matter, a fact that we identify with a cosmological analogue of the vDVZ discontinuity. The main findings of this work are discussed in Section \ref{sec6}, where the cosmological Vainshtein screening mechanism is invoked to resolve the cosmological vDVZ discontinuity. Finally, in section \ref{sec-conclu} brief conclusions are given. In this paper we use the units where $8\pi G_N=c=h=1$.

%%%%%%%%%%%%%%%%%%%%%%%%%%%%%%%%%%%%%%%%%%%%%%%%%%

\section{Cosmological equations}\label{sec2}

In what follows we adopt a flat FRW background metric with line element: $ds^2=-dt^2+a^2(t)\delta_{ik}dx^i dx^k.$ The cosmological field equations resulting from the action \eqref{action} then read:

\bea &&3H^2=\rho_m+\rho_\text{eff},\label{fried}\\
&&-2\dot H=\rho_m+p_m+\rho_\text{eff}+p_\text{eff},\label{ray}\\
&&\dot\rho_m+3H(\rho_m+p_m)=0,\label{ecm}\\
&&\left(1+2\sigma_{,\phi}\dot\phi^2-6\sigma H\dot\phi\right)\ddot\phi+3H\dot\phi+\left(\frac{1}{2}\sigma_{,\phi\phi}\dot\phi^2-3\sigma\dot H-9\sigma H^2\right)\dot\phi^2=-V_{,\phi},\label{ecc}\eea where $\rho_m\propto a^{-3(\omega_m+1)}$ is the energy density of the background matter fluid with equation of state $p_m=\omega_m\rho_m$ ($p_m$ is the barotropic pressure and $\omega_m$ is a constant parameter). It is important to mention here that equations \eqref{fried}-\eqref{ecc} are not independent of each other due to the Bianchi identities. We can write the effective energy density and the effective (parametric) pressure of the galileon field in the following way: 

\bea &&\rho_\text{eff}=\frac{\dot\phi^2}{2}\left(1+\sigma_{,\phi}\dot\phi^2-6\sigma H \dot\phi\right)+V,\nonumber\\
&&p_\text{eff}=\frac{\dot\phi^2}{2}\left(1+\sigma_{,\phi}\dot\phi^2+2\sigma\ddot\phi\right)-V.\label{par}\eea
 
In this paper, for simplicity, we focus in the constant (positive) galileon coupling: $\sigma=\sigma_0$ ($\sigma_0>0$). Under this choice the motion equations appreciably simplify. The resulting cosmological field equations are \eqref{fried} and \eqref{ray} with

\bea &&\rho_\text{eff}=\frac{\dot\phi^2}{2}\left(1-6\sigma_0H\dot\phi\right)+V,\nonumber\\
&&p_\text{eff}=\frac{\dot\phi^2}{2}\left(1+2\sigma_0\ddot\phi\right)-V,\label{rho-p}\eea plus the equation of motion of the galileon:

\bea &&\left(1-6\sigma_0H\dot\phi\right)\ddot\phi+3H\dot\phi-3\sigma_0H^2\left(3+\frac{\dot H}{H^2}\right)\dot\phi^2=-V_{,\phi}.\label{kg-eq}\eea

The effective equation of state parameter (EOS) of the dark energy $\omega_\text{eff}$ and the deceleration parameter $q$ are given by:

\bea \omega_\text{eff}:=\frac{p_\text{eff}}{\rho_\text{eff}}=\frac{\frac{\dot\phi^2}{2}\left(1+2\sigma_0\ddot\phi\right)-V}{\frac{\dot\phi^2}{2}\left(1-6\sigma_0H\dot\phi\right)+V},\label{eff-ode}\eea and, 

\bea &&q:=-1-\frac{\dot H}{H^2}=-1+\frac{3}{2}(\omega_m+1)\Omega_m+\frac{3}{2}(\omega_\phi+1)\Omega_\phi+\frac{\sigma_0\dot\phi^2}{2H^2}(\ddot\phi-3H\dot\phi),\label{q}\eea respectively. In the latter equation $\Omega_c\equiv\rho_c/3H^2$ stands for the dimensionless energy density parameter of the '$c$'-component of the background cosmic fluid, and $$\omega_\phi\equiv\frac{\dot\phi^2-2V}{\dot\phi^2+2V}.$$

\subsection{Main (simplifying) assumptions}

Combining the first equation in \eqref{rho-p} with \eqref{fried}, for the Hubble rate one gets

\bea H_\pm=-\frac{\sigma_0\dot\phi^3}{2}\pm\sqrt{\frac{\sigma_0^2\dot\phi^6}{4}+\frac{1}{3}\left(\rho_m+\rho_\phi\right)},\label{h-eq}\eea where we have taken into account the standard definition of the scalar field's energy density $\rho_\phi:=\dot\phi^2/2+V$. The '$\pm$'
signs in the right-hand side (RHS) of \eqref{h-eq} are for two possible branches of the cosmological evolution in the model: Assuming non-negative $\sigma_0$ the '$+$' branch represents universes that expand for ever, while the '$-$' branch is for ever contracting universes. In this paper we shall focus in expanding cosmology exclusively so that in what follows we shall consider only the '+' branch of \eqref{h-eq}. This choice will not affect neither the generality of our analysis, nor the validity of the results of the present research. Summarizing, the main simplifying assumptions of the present paper are:

\begin{itemize}

\item We focus in non-negative (constant) non linear coupling $\sigma_0\geq 0$, 

\item ever expanding universes: $H>0$ will be the subject of this paper, and

\item only non-negative potentials $V\geq 0$ drive viable cosmological behavior.

\end{itemize} In section \ref{sec6} we shall discuss on the domain of validity of the above assumptions.

%%%%%%%%%%%%%%%%%%%%%%%%%%%%%%%%%%%%%%%%%%%%%%%%%%%%%%%%%%%%%%%%%%%%%%%%%%%%%%%%%%%

\section{Self-interaction potentials beyond the exponential one}\label{sec-bey-exp}

There are both theoretical and observational motivations for studying cosmological models with more complicate potentials than the exponential one. So far a wide variety of scalar field dark energy models have been proposed to explain, among other things, the early inflation era, the large scale structure and the late time cosmic acceleration. In this work we will focus in four self-interacting potentials with interesting cosmological consequences: i) the power-law, ii) a potential for the tachyon field, iii) a supergravity motivated potential, and iv) the double exponential potential.

\begin{enumerate}

\item The original quintessence model is described by the power-law potential: 

\bea V(\phi)=\frac{M^{4+\alpha}}{\phi^{\alpha}},\label{p-law-pot}\eea where $\alpha$ is a positive number and $M$ is a constant with units of mass. Such kind of potential is known under the name of Ratra-Peebles and can be found in models of supersymmetric QCD. These are extensively studied because of their late-time behavior which allows for a solution -- or at least an alleviation -- of the initial conditions problem \cite{axel}. Models with $\alpha>0$ are more interesting for the dark energy phenomenology, while the case $\alpha <0$ is largely studied in the investigation of the early-time inflation. One of the problems of the corresponding quintessence models is that the quintessence must be coupled to ordinary matter, leading to long range forces and to time dependence of the constants of nature.

\item It has been suggested that tachyon condensates may have interesting cosmological dynamics in certain class of string theories. In the case that the tachyon field starts to roll down slowly the potential 

\bea V(\phi)=V_0[\cosh(\alpha\phi/M_\text{Pl})]^{-k},\label{cosh-pot}\eea where $V_0$ is a constant and, as above, $M_\text{Pl}$ is the Planck mass, an early universe dominated by this field evolves smoothly from a phase of accelerated expansion to an era dominated by a non-relativistic fluid.  Furthermore, depending on the parameters of the potential the tachyon field can also act as a source of dark energy giving a power-law expansion $a(t) \propto a^t$ \cite{Pad}. This is why we consider the above potential in the present investigation.

\item Models of quintessence in supergravity have been constructed leading to interesting phenomenological consequences such as low values of the equation of state paramter $\omega$.  Furthermore, SUSY search in different forms is an open studied subject of particle physics research due to its theoretical appeal and phenomenological implications. However, to derive a quintessence model from string theory it is necessary to satisfy a no go theorem which states that can not be a scalar field with positive potential \cite{Maldacena}. In order to avoid this problem in \cite{Brax} it was imposed the condition that the expectation value of the superpotential vanish and this leads to the potential 

\bea V(\phi)=\frac{M^{4+\alpha}}{\phi^\alpha}\,e^{\phi^2/2M_\text{Pl}}.\label{sugra-pot}\eea

\item The double exponential potential: 

\bea V(\phi)=V_1\,e^{-\mu_1\phi/M_\text{Pl}}+V_2\,e^{-\mu_2\phi/M_\text{Pl}},\label{dexp-pot}\eea has been used in order to obtain acceptable solutions for a wider range of initial energy densities \cite{Rubano}. In \cite{Cora} the authors have used the Supernova data (excluding the recent data at z= 1.7) and measurements of the position of the acoustic peaks of the CMBR spectra to constrain a general class of potentials. They have argued that in order to have the equation of state parameter $\omega_\phi\sim -1$, the quintessence field has to evolve in a very flat region of the potential, and such behavior can be obtained from the double exponential potential.
 
\end{enumerate}

In the rest of this paper we shall focus in the study of the phase space dynamics of the cubic galileon models based in the action \eqref{action} with the above listed potentials.

%%%%%%%%%%%%%%%%%%%%%%%%%%%%%%%%%%%%%%%%%%%%%%%%%%

\section{The dynamical system}\label{sec3}

Our aim here will be to trade the very complex system of second order equations \eqref{fried}, \eqref{ray}, \eqref{ecm}, \eqref{rho-p} and \eqref{kg-eq} by a system of autonomous ordinary differential equations (ODE) on certain variables of the equivalent phase space. For this purpose one has to choose adequate variables of the state space. In general there are many possible ways to achieve this task, nevertheless the most common one is to consider the expansion normalized variables \cite{wands}:

\bea x_s:=\frac{\dot\phi}{\sqrt{6} H},\;y_s:=\frac{\sqrt V}{\sqrt{3}H}.\label{xy-var}\eea In terms of these variables the evolution equations can be written as

\bea &&x_s'=\frac{\ddot\phi}{\sqrt{6} H^2}-x_s\frac{\dot H}{H^2},\nonumber\\
&&y_s'=-y_s\left[\sqrt{\frac{3}{2}}\lambda x_s+\frac{\dot H}{H^2}\right],\label{ode-xy-s}\eea where the prime denotes the derivative with respect to the time variable $\tau\equiv\ln a$, and we have defined the function $\lambda=\lambda(\phi)$, in the following way:

\bea \lambda:=-\frac{V_{,\phi}}{V}.\label{lamb}\eea The dynamical equation for this latter function reads
 
\bea \lambda'=-\sqrt{6}\lambda^2 x_s(\Gamma-1),\label{gam}\eea where the function $\Gamma=\Gamma(\phi)$ is defined as follows:

\bea \Gamma:=\frac{V V_{,\phi\phi}}{(V_{,\phi})^2}.\label{Gamma}\eea 

Here we should emphasize that the above dynamical variables fail to close the system of equations since, in general, the parameter $\Gamma$ is itself a dynamical variable so that an additional evolution equation is required. However, since both $\lambda$ and $\Gamma$ are functions of $\phi$, it is possible -- in principle -- to relate one to another for given self-interaction potentials \cite{Fang} (see also \cite{quiros}). In other words, provided that the function $\lambda(\phi)$ is invertible, we can write $\phi(\lambda)$ and then $\Gamma$ can be given as a function of $\lambda$. By defining a new function

\bea f(\lambda):=\lambda^2[\Gamma(\lambda)-1],\label{f-eq}\eea the dynamical equation for $\lambda$ takes the simpler form

\bea \lambda'=-\sqrt{6}x_s f(\lambda).\label{ode-l}\eea

In order to illustrate the working idea behind the approach let us take, as an example, the well-known exponential potential (this case was studied in \cite{rtgui}). For this choice of the potential the function $\Gamma(\lambda)=1$ and $\lambda$ is not a dynamical variable. In consequence the autonomous system \eqref{ode-xy-s}, \eqref{ode-l} reduces to \eqref{ode-xy-s}. In general, for more complicated potentials, the function $f(\lambda)$ vanishes only at equilibrium configurations and the dynamical system is composed of the following autonomous ODE-s:

\bea &&x_s'=\frac{\ddot\phi}{6 H^2}-x_s\frac{\dot H}{H^2}, \nonumber \\
&&y_s'=-y_s\left[\sqrt\frac{3}{2}\lambda x_s+\frac{\dot{H}}{H^2}\right],\nonumber\\
&&\lambda'=-\sqrt{6}x_s f(\lambda).\label{ode-syst}\eea
		
Before we consider specific functions $\Gamma(\lambda)$ -- i. e. specific potentials $V(\phi)$ -- we shall extract as much information from the dynamical system \eqref{ode-syst} as we can, leaving the parameter $\Gamma$ as an arbitrary function of $\lambda$. Using the set of variables \eqref{xy-var}, the Friedmann constraint equation \eqref{fried} becomes

\bea \Omega_m=1-x_s^2-y_s^2+6\sqrt{6}\,x^3_s H^2 \sigma_0.\label{fried-c}\eea We want to underline that in the limit $H^2 \sigma_0\ll 1$ the standard quintessence scenario is recovered from the present model (see below for further discussion on this issue).

As seen from \eqref{fried-c}, one needs yet another variable to account for the factor $H^2\sigma_0$. Furthermore, due to the positive sign of the fourth term in the right-hand side (RHS) of \eqref{fried-c}, given $x_s\geq 0$, the variables $x_s$ and $y_s$ can take arbitrary large values, while $0\leq\Omega_m\leq 1$.

\subsection{Finite-size phase space}

It is desirable to work in a finite-size phase space, hence it would be appropriate to choose the following bounded variables 

\bea x_\pm=\frac{1}{x_s\pm1},\;y=\frac{1}{y_s+1},\;z=\frac{1}{H^2\sigma_0+1},\;v=\frac{1}{\lambda+1},\label{n-var}\eea where $x_+$ is for non-negative $x_s$ ($\dot\phi\geq 0$), while $x_-$ is for non-positive $x_s$ ($\dot\phi\leq 0$), besides, $0\leq x_+\leq 1$ ($-1\leq x_-\leq 0$), $0\leq y\leq 1$, $0\leq z\leq 1$, and $0 \leq v \leq 1$. As already stated in section \ref{sec2}, here we are assuming that only expanding
cosmologies arise: $H>0$ ($y_s>0$), and that the galileon is a monotonically growing function of the cosmic time: $\dot\phi>0$, so that along orbits of the phase space $x_s$ does not flip sign. The choice of coordinates in \eqref{n-var} is specially useful in those cases where $x_s=0$, and $y_s=0$ are separatrices FIG. \ref{fig-1}. As we will show below, this is, precisely, the case for the vacuum of the generalized galileon model.

The definition of the coordinate $z$ in \eqref{n-var} deserves a few more words. Due to the definition $z$ asymptotically approaches unity whenever $\sigma_0 H^2\rightarrow 0$. It is clear that the coordinates \eqref{n-var} do not cover the situation where $H=0$, in which case the variables $x_s$ and $y_s$ are ill-defined; or when the Hubble factor changes sign, in which case the arrow of time defined by $f'=1/H\dot f$ is reverted. Hence, in this paper the phase space points at $z=1$ will correspond to standard quintessence model points,\footnote{An exception are the equilibrium points $P_{7v}$ and $P_{8v}$ in section \ref{sec5} -- corresponding to the vacuum galileon case -- that are genuinely linked with the cubic self-interaction ($\sigma_0\neq 0$). However, the mentioned critical points are unstable nodes (local past attractors) without impact in the late-time cosmic dynamics.} i. e., those for which the cubic self-interaction may be ignored: $\sigma_0\ll H^{-2}$.

The analysis of contracting and of bouncing solutions (in case these existed), or of any configuration with $H=0$ or $H<0$, is beyond the scope of the present research and can be solved in a forthcoming paper. Nevertheless, if we were interested in the complete analysis of the situations where $H=0$, or where $H$ changes sign, we cannot use the $H$-normalization, but we have to adopt another normalization instead. One way is to define alternative dynamical variables and an alternative time derivative that is well-defined when $H=0$ (e. g., similar to the variables used in \cite{Karpathopoulos:2017arc, Giacomini:2017yuk}):

\bea X_s=\frac{\sqrt{\sigma_0} \dot \phi}{\sqrt{6} \sqrt{1+\sigma_0 H^2}},\;Y_s=\frac{\sqrt{\sigma_0}\sqrt{V}}{\sqrt{3}\sqrt{1+\sigma_0 H^2}},\;Z=\frac{\sqrt{\sigma_0} H}{\sqrt{1+\sigma_0 H^2}},\label{eq_27-new}\eea which are related with the original variables $x_s$, $y_s$ and $z$ through $X_s=x_s Z$, $Y_s=y_s Z$, $z=1-Z^2$, and the time variable $\bar{\tau}$ such that $$\frac{d f}{d\bar{\tau}}\equiv\frac{\sqrt{\sigma_0}}{\sqrt{1+\sigma_0 H^2}}\dot f=\frac{Z}{H}\dot f=Z\frac{df}{d\tau}.$$ By definition the sign of $H$ is the same as the sign of $Z$, such that $Z<0$ corresponds to contracting cosmologies, $Z>0$ corresponds to expanding cosmologies, and $Z=0$ corresponds to $H=0$, and in this case the variables and the time derivative are well-defined as $H=0$. Finally, using the aforementioned compactification procedure we get

\bea \frac{dZ}{d\bar{\tau}}|_{Z=0}=3\left[X_s^2 \left(w_m-6 \lambda  Y_s^2-1\right)+(w_m+1) Y_s^2\right].\nonumber\eea This derivative has not definite sign.\footnote{In this case, for instance, the problem of finding a value $a_r$ where an static universe ($H=0$, $\dot H=0$) is located, is equivalent to find the value of $\bar{\tau}$ such that $3\left[X_s(\bar{\tau})^2 \left(w_m-6 \lambda Y_s(\bar{\tau})^2-1\right)+(w_m+1)Y_s(\bar{\tau})^2\right]=0$.} Using the variables \eqref{eq_27-new} and the time variable $\bar{\tau}$, one can solve issues that cannot be properly addressed using $x_s$, $y_s$, $z$ (or using the variables \eqref{n-var}), like the change from contraction to expansion and vice versa, and also the problem of finding bouncing cosmologies, etc. 

In the next sections we shall focus on expanding cosmologies and we will discuss whether the late-time dynamics observed in exponential self-interaction potentials also exists in other potentials of cosmological interest. As a remarkable result we will put into context our main claim that the inclusion of matter fields may screen the phantom-like effects of the galileon, an effect that can be interpreted as a kind of cosmological vDVZ-type discontinuity.

\subsection{Cosmological parameters}

In order to perform the present analysis it will helpful to compute the potential related parameter $\Gamma$ and other helpful functions of the slow-roll parameter such as 

\bea \left.\Gamma_*\equiv\frac{d(\Gamma-1)}{d\lambda}\right|_{\lambda_*},\label{Gamma-aster}\eea that are evaluated at the roots $\lambda=\lambda_*$ of $f(\lambda)=0$ in the third equation in \eqref{ode-syst}. For the potentials of interest in the present investigation we have:

\begin{itemize}

\item Power-law potential \eqref{p-law-pot}, 

\bea V(\phi)=V_0 \phi^{-p}\;\Rightarrow\;f(\lambda)=\frac{\lambda^2}{p},\;\lambda_*=0,\;\Gamma_*=0.\label{pwl-pot}\eea

\item Cosh potential \eqref{cosh-pot},

\bea &&V(\phi)=V_0[\cosh(j\,\phi)]^{-k}\;\Rightarrow\;f(\lambda)=\frac{\lambda^2-k^2j^2}{k},\;\lambda_*=kj,\;\Gamma_*=\frac{2j}{k^2}.\eea

\item Combined power-law-exponential \eqref{sugra-pot}, 

\bea &&V(\phi)=V_0\phi^{-m}\,e^{-n\phi}\;\Rightarrow\;f(\lambda)=\frac{(\lambda-n)^2}{m},\;\lambda_*=n,\;\Gamma_*=0.\eea

\item Double exponential \eqref{dexp-pot},

\bea &&V(\phi)=V_1\,e^{-r\phi}+V_2\,e^{-s\phi}\;\Rightarrow\;f(\lambda)=-(\lambda-r)(\lambda-s),\;\lambda_*=\{r,\,s\},\;\Gamma_*=\left\{\frac{s-r}{r^2},\,\frac{r-s}{s^2}\right\}.\eea

\end{itemize}

%%%%%%%%%%%%%%%%%%%%%%%%%%%%%%%%%%%%%%%%%%%%%%%%%%%%%%%%%%%%%%%%%%%

\section{Cubic galileon cosmology with matter}\label{sec4}

For simplicity we assume that the matter content of the universe is described by pressureless dust ($p_m=0$). The dynamical system corresponding to the cosmological field equations \eqref{fried}, \eqref{ray}, \eqref{ecm}, \eqref{rho-p} and \eqref{kg-eq}, in terms of the bounded variables \eqref{n-var}, is given by:

\bea &&x'_\pm=-\frac{x_\pm^2}{\sqrt 6}\left(\frac{\ddot\phi}{H^2}\right)_\pm+x_\pm(1\mp x_\pm)\left(\frac{\dot H}{H^2}\right)_\pm,\nonumber\\
&&y'=y(1-y)\left[\sqrt\frac{3}{2}\left(\frac{1-v}{v}\right)\left(\frac{1\mp x_\pm}{x_\pm}\right)+\left(\frac{\dot H}{H^2}\right)_\pm\right],\nonumber\\
&&z'=2z(z-1)\left(\frac{1\mp x_\pm}{x_\pm}\right),\nonumber\\
&&v'=\sqrt{6}\left(\frac{1\mp x_\pm}{x_\pm}\right)\,F(v),\label{ode-mat}\eea where $F(v)\equiv v^2f(v)$, and

\bea &&\left(\frac{\dot H}{H^2}\right)_\pm=\frac{3x_\pm(1\mp x_\pm)\left[x_\pm(1\mp x_\pm)(1-y)^2(1\mp v)-\sqrt{6}\Theta_\pm(2)v\right]Q-6(1\mp x_\pm)^4 y^2 Q^2 v-\frac{9}{2}x_\pm^2\Theta_\pm(1)v}{y^2 v \left[3x_\pm^4+2\sqrt{6}x_\pm^3(1\mp x_\pm)Q+2(1\mp x_\pm)^4Q^2\right]},\nonumber\\
&&\left(\frac{\ddot\phi}{H^2}\right)_\pm=\frac{\left\{9x_\pm^4 (1-y)^2(1-v)-9\sqrt{6} x_\pm^3(1\mp x_\pm)y^2-9(1\mp x_\pm)^2\Delta_\pm Q\right\}}{y^2v\left[3x_\pm^4+2\sqrt{6}x_\pm^3(1\mp x_\pm)Q+2(1\mp x_\pm)^4Q^2\right]}.\label{def-s}\eea We have also defined $Q\equiv-9\sigma_0 H^2=9(z-1)/z$, and the following functions:

\bea \Theta_\pm(a):=a(1\mp x_\pm)^2y^2-x^2_\pm(1-2y),\;\Delta_\pm:=x_\pm^2(1-y)^2-(1\mp2 x_\pm)y^2.\label{defs}\eea

In the autonomous system of ODE \eqref{ode-mat} the '$\pm$' signs account for two different branches of the dynamical system so that, as a matter of fact, one has two different dynamical systems. First we will analyze the system of equations for a general galileon cubic model regardless of the form of the potential and then we will focus in the specific potentials \eqref{p-law-pot}-\eqref{dexp-pot}.

%%%%%%%%%%%%%%%%%%%%%%%%%%%%%%%%%%%%%%%%%%%%%%%%%%

\begin{table*}[tbh]\centering
\begin{tabular}{||c||c|c|c|c|c|c|c|c|}
\hline\hline
Crit. Point  &  $x_{\pm}$  & $y$  &  $z$ & $v$  &   Existence & $\Omega_m$ & $\omega_\text{eff}$ & $q$\\
\hline\hline
$P_1^{\pm}$ &  $\pm 1$ & $1$  &  $0$  & $v$  & always  &  $1$ &  undefined  &  $1/2$ \\
\hline
$P_2^{\pm}$ & $\pm 1$  & $1$  &  $1$  &  $v$  & always & $1$  &  undefined &  $1/2$   \\
\hline
$P_3^{\pm}$ & $\pm 1$ &  $1/2$  & $1$ & $1$  & $\lambda_*=0$ &  $0$  & $-1$  & $-1$ \\
\hline
$P_4^{\pm}$ & $\pm 1/2$  &  $1$  &  $1$ & $\frac{1}{\lambda_*+1}$ & always & $0$ & $1$ & $2$ \\
\hline
$P_5^{\pm}$ &$\frac{\sqrt 6}{\lambda_*\pm\sqrt{6}}$&$\frac{\sqrt 6}{\sqrt{6}+\sqrt{6-\lambda_*^2}}$&$1$&$\frac{1}{\lambda_*+1}$&$\lambda_*^2<6$&$0$&$-1 +\frac{\lambda_*^2}{3}$&$-1+\frac{\lambda_*^2}{2}$\\
\hline
$P_6^{\pm}$ &$\frac{2\lambda_*}{\sqrt{6}\pm 2\lambda_*}$&$\frac{2\lambda_*}{\sqrt{6}+2\lambda_*}$&$1$&$\frac{1}{\lambda_*+1}$&$\lambda_*^2>3$&$1-\frac{3}{\lambda_*^2}$&$0$&$1/2$ \\
\hline\hline\end{tabular}
\caption{The physically meaningful critical points of the autonomous system \eqref{ode-mat} together with their existence conditions, the corresponding values of the dimensionless matter density parameter $\Omega_m$, of the effective EOS $\omega_\text{eff}$ and of the deceleration parameter $q$.}\label{tab-1}
\end{table*}

%%%%%%%%%%%%%%%%%%%%%%%%%%%%%%%%%%%%%%%%%%%%%%%%%%

\begin{table*}[tbh]\centering
\begin{tabular}{||c||c|c|c|c|c|c|c|}
\hline\hline
Crit. Point  & $\lambda_1$ & $\lambda_2$ & $\lambda_3$ & $\lambda_4$ & Stability \\
\hline\hline
$P_1^{\pm}$ & $3$ & $-3/2$ & $3/2$ & $0$ & saddle \\
\hline
$P_2^{\pm}$ & $-3$ & $-3/2$ & $3/2$ & $0$ & saddle \\
\hline
$P_3^{\pm}$ & $-3$ & $0$ & $-\frac{3}{2}+\beta$ & $-\frac{3}{2}-\beta$ & stable if $f(0)\leq 0$ \\
            &      &     &                      &                      &  saddle  if  $f(0)>0$ \\
\hline
$P_4^{\pm}$ & $-6$ & $3$ & $3\mp\sqrt\frac{3}{2}\lambda_*$ & $\mp\sqrt{6}\lambda_*^2\Gamma_*$ & saddle \\
\hline
$P_5^{\pm}$ & $-\lambda_*^2$ & $-3+\frac{\lambda_*^2}{2}$ & $-3+\lambda_*^2$ & $-\lambda_*^3\Gamma_*$ & stable if $\lambda_*^2<3$ and $\lambda_*\Gamma_*>0$ \\
            &                &                            &                  &                        &  saddle if $\lambda_*^2>3$ or $\lambda_*\Gamma_*<0$ \\ 
\hline
$P_6^{\pm}$ & $-3$ & $-\frac{3}{4}\pm\alpha$ & $-\frac{3}{4}\pm\alpha$ & $-3\lambda_*\Gamma_*$ & stable if $\lambda_*\Gamma_*>0$ \\
            &      &                         &                         &                       & saddle otherwise \\
\hline\hline\end{tabular}
\caption{The physically meaningful critical points of the autonomous system \eqref{ode-mat}, together with the eigenvalues of the corresponding linearization matrices, and their consequent stability properties. Here we have defined $\alpha\equiv\frac{3}{4\lambda_*}\sqrt{24-7\lambda_*^2}$ and $\beta\equiv\frac{\sqrt 3}{2}\sqrt{3-4f(0)}$.} \label{tab-2}\end{table*}

%%%%%%%%%%%%%%%%%%%%%%%%%%%%%%%%%%%%%%%%%%%%%%%%

\subsection{Critical points and stability}

We start by recalling the properties of the critical points $({x_{\pm *},y_*,z_*,v_*})$, i. e., those for which $x_{\pm}'=y'=z'=v'=0$. The physically meaningful critical points of the dynamical system \eqref{ode-mat}, together with their existence conditions, the values of the dark matter density parameter $\Omega_m$, of the effective (galileon) EOS $\omega_\text{eff}$ and of the deceleration parameter $q$, are shown in TAB. \ref{tab-1}, while in TAB. \ref{tab-2} we display the eigenvalues of their linearization matrices and we summarize the corresponding stability properties. 

These tables reflect the fact that, but for the matter-dominated big bang $P_1^\pm$: $H\rightarrow\infty$, which is independent of the value $v$ and, hence, of the functional form of the self-interaction potential, the present galileon model does not differ too much from the standard quintessence. Actually, since for the remaining equilibrium points in TAB. \ref{tab-1}: $z=1$, and since -- as we have already clearly established -- the phase space coordinates \eqref{n-var} cannot cover cases where $H=0$, this means that these critical pints are to be associated with quintessence behavior (vanishing of the cubic self-interaction term: $\sigma_0\ll H^{-2}$, $H\neq 0$).\footnote{From the equation \eqref{fried-c}, it is seen that when the third term in the right hand side is negligible, i.e., when  $\sigma_0 H^2\ll 1$, the Friedmann constraint $\Omega=1-x_s^2-y_s^2$ is closely recovered, which is the relation that arises in the standard quintessence model \cite{wands} whenever $H\neq 0$. We emphasize that this argument is about the relative values of $\sigma_0$ and $H^2$, and does not necessarily mean that one of them exactly vanishes.} Despite that these equilibrium points are essentially the same as those found in TAB.1 of \cite{Fang}, for sake of completeness, below we list their main properties.

\begin{enumerate}

\item The dark matter dominated (decelerating) solution $P_2^\pm$: $3H^2=\rho_m$, which is independent of the form of the potential. It is associated with a saddle critical manifold along the $v$ direction in the phase space. This is why the eigenvalue of the corresponding linearization matrix that is associated with the eigenvector along the $v$-direction, vanishes: $\lambda_4=0$. In the standard normalized variables this point corresponds to $x_s=0$, $y_s=0$, and since the energy density of the galileon vanishes, the effective EOS parameter $\omega_\text{eff}$ is undetermined. 

\item The de Sitter (accelerated expansion) solution $P_3^\pm$: $$H=\sqrt{V_0/3}\;\;(\dot\phi=0\Rightarrow V=V_0),$$ corresponds to nothing but a cosmological constant. As in the standard quintessence cosmology this point exists only for the constant self-interaction potentials, or for potentials that asymptote to a non-vanishing constant. As we can see one of the eigenvalues of the Jacobian vanishes implying that this point is a non-hyperbolic critical point. Since the real parts of all of the remaining eigenvalues are negative, in order to determine the stability we must either to be able to find a Liapunov function, to apply the centre manifold theorem \cite{center, center2, centre} or -- as we do in the present paper -- to resort to the numerical investigation. This singular point is quite important, since it represents a local attractor of the dynamical system (hence it may represent the late-time dynamics of the universe), and possesses the parameters $\omega_\text{eff}=-1$ and $q=-1$ that are compatible with observations.

\item The stiff-matter solutions $P_4^\pm$: $3H^2=\dot\phi^2/2$, exist for all values of $\lambda_*$, so that their phenomenological properties remain the same independently of the potential. From TAB. \ref{tab-2} we see that one eigenvalue of the corresponding Jacobian matrix: $\lambda_4$, may have a vanishing real part if either $\lambda_*=0$ or $\Gamma_*=0$. This is to be expected since, unless one specifies the functional form of the self-interaction potential, this solution represents a linear manifold along the $v$-direction. Once one specifies the form of the potential, a specific value $v=1/(\lambda_*+1)$ is picked up, implying that points $P_4^\pm$ may be isolated critical points. Since the real parts of the remaining eigenvalues have different sign, we conclude that $P_4^\pm$ represent a saddle node.

\item The points $P_5^\pm$: $$3H^2=\frac{\dot\phi^2}{2}+V,$$ are related with the quintessence-dominated solution. Their existence depends on the concrete form of the self-interacting potential, and is given by the bound: $\lambda_*^2\leq 6$. These points attract the universe at late-times if $\lambda_*^2<3$ and $\lambda_*\Gamma_*>0$. Otherwise, if either $3<\lambda_*^2<6$, or $\lambda_*\Gamma_*<0$ (or both), $P_5^\pm$ are saddle critical points and, correspondingly, these represent a transient stage of the cosmic evolution.

\item The matter-scaling solutions $P_6^\pm$: $$\frac{\Omega_m}{\Omega_\phi}=\frac{\lambda_*^2}{3}-1,$$ where we have taken into account the fact that $\Omega_\text{eff}=\Omega_\phi$ since $\sigma_0=0$, represent decelerating solutions where the quintessence tracks the dark matter behavior $\omega_\text{eff}=\omega_\phi=\omega_m=0$. Their existence is related to the concrete form of the self-interacting potential and is given by the bound: $\lambda_*^2>3$. These solutions represent always saddle points in the phase-space since the non-vanishing (real parts of the) eigenvalues of the linearization matrix have opposite sing. In other words, the matter-scaling solutions can represent, at most, transient stages of the cosmic evolution.

\end{enumerate} As already said, the above results are essentially the same obtained in \cite{Fang} by means of a bit different procedure.

\begin{figure*}\centering
\includegraphics[width=5cm]{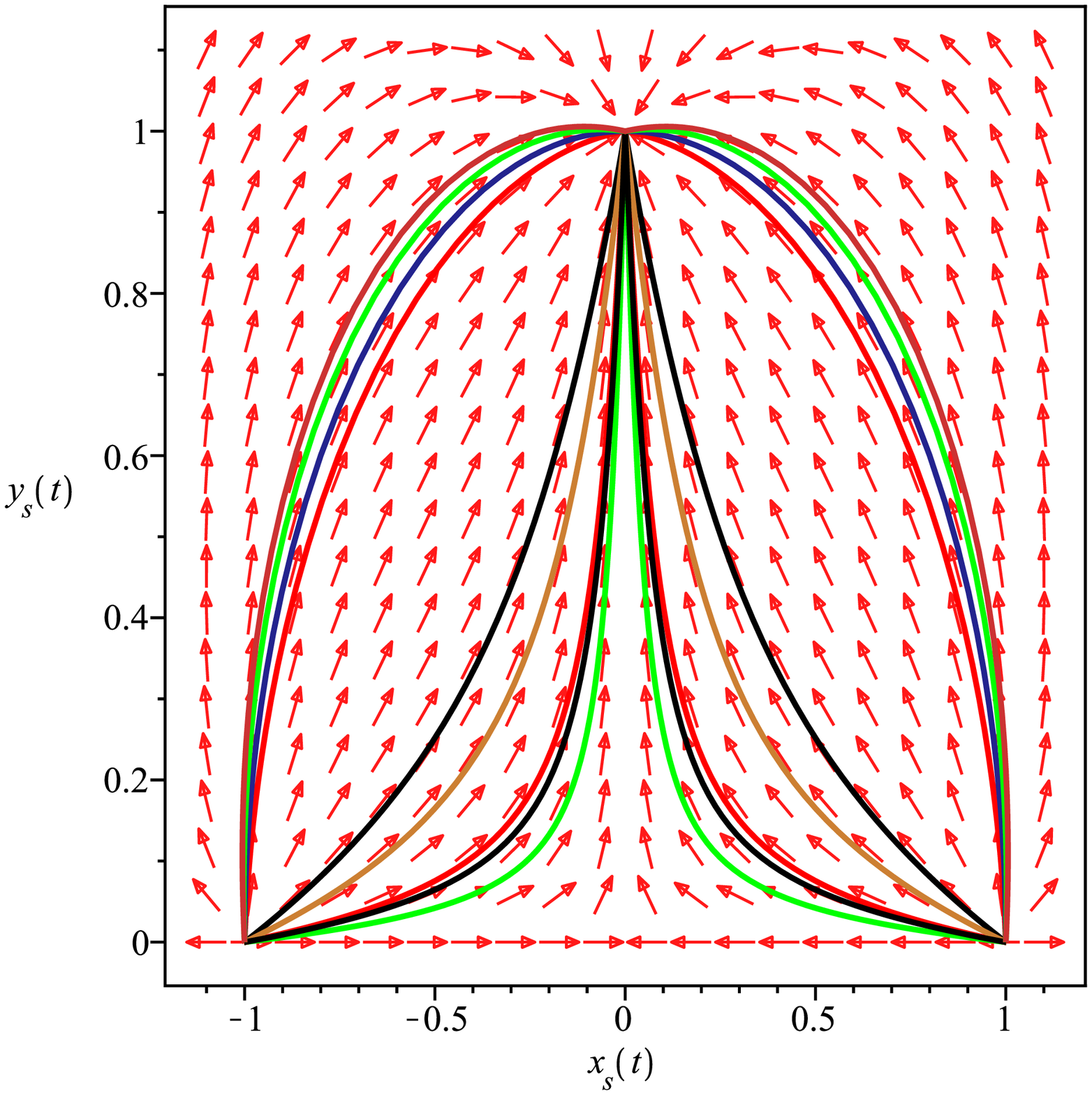}
\includegraphics[width=5cm]{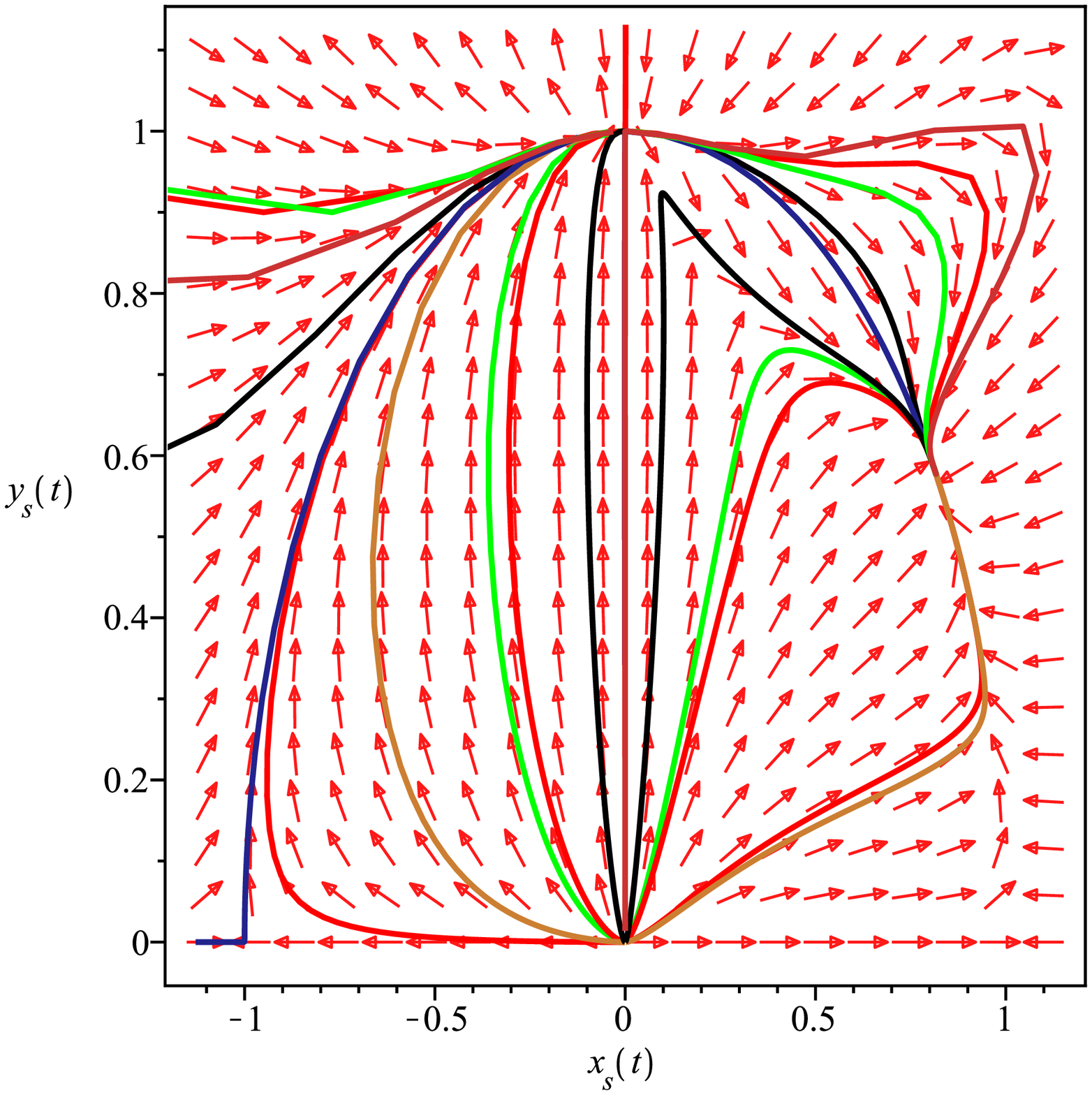}
\includegraphics[width=5cm]{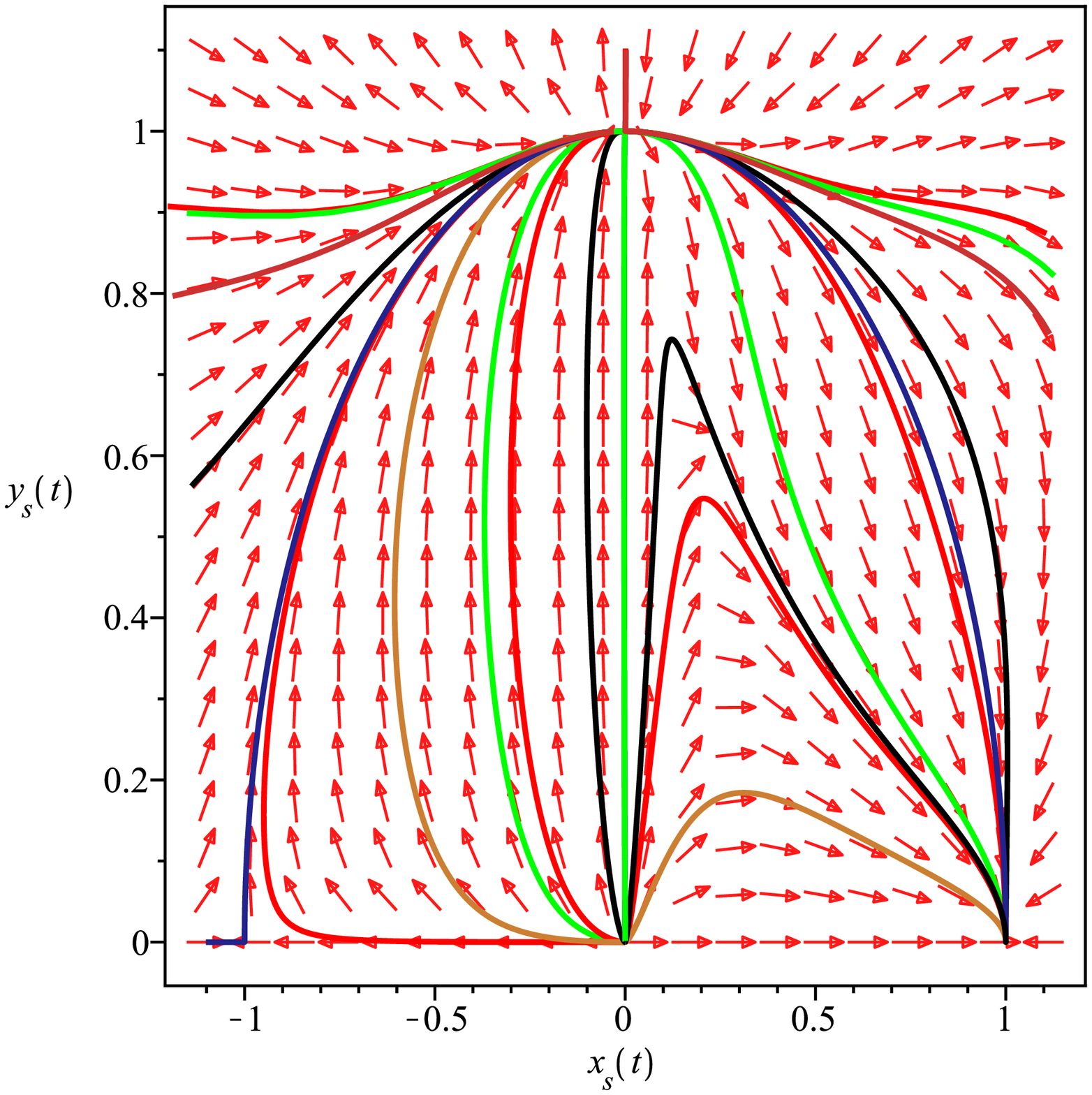}\vspace{0.7cm}
\caption{Phase portrait of the dynamical system \eqref{asode-vac} for different values of the parameter $\lambda$. From left to the right $\lambda=0$, $\lambda=2$ and $\lambda=5$. It is seen that, thanks to the galileon coupling $\sigma_0$, the orbits are not confined within the semi-disk $x_s^2+y_s^2\leq 1$, but these may also evolve outside it.}\label{fig-1}
\end{figure*}

%%%%%%%%%%%%%%%%%%%%%%%%%%%%%%%%%%%%%%%%%%%%%%%%%%%%%%

\section{Cubic galileon vacuum}\label{sec5}

Apparently, the simplest case we can deal with is when the cosmic background is the vacuum ($\Omega_m=0$). One may naively expect that there can be no new interesting dynamics in the vacuum case with respect to the results of the previous, more general case, where the background matter is considered. Quite the contrary: as we shall show in this section, a new asymptotics arises in the vacuum case when compared with the dynamics of background matter case.

After setting $\Omega_m=0$, in terms of the standard normalized variables \eqref{n-var} the Friedmann constraint \eqref{fried} amounts to a relationship between $x_s$, $y_s$ and $z$:

\bea z=\frac{6\sqrt{6}x_s^3}{6\sqrt{6}x_s^3+x_s^2+y_s^2-1},\label{fr-g0-exp}\eea so that one of these variables, say $z$ is redundant, and one ends up with a system of ODE

\bea &&x'_s=\frac{1}{\sqrt 6}\frac{\ddot\phi}{H^2}-x_s\frac{\dot H}{H^2},\nonumber\\
&&y'_s=-y_s\left(\sqrt\frac{3}{2}\,\lambda x_s+\frac{\dot H}{H^2}\right),\nonumber \\
&&\lambda'=-\sqrt{6}x_s\lambda^2 f(\lambda),\label{asode-vac}\eea where

\bea &&\frac{\dot H}{H^2}=-\frac{6(1-y_s^2)(1+x_s^2-y_s^2)}{4(1-y_s^2)+(x_s^2+y_s^2-1)^2}+\frac{3(x_s^2+y_s^2-1)(1+x_s^2-y_s^2-\sqrt{2/3}\,\lambda\,x_sy_s^2)}{4(1-y_s^2)+(x_s^2+y_s^2-1)^2},\nonumber\\
&&\frac{\ddot\phi}{H^2}=-\frac{3\sqrt{6}\,x_s(x_s^2+y_s^2-1)(1+x_s^2-y_s^2)}{4(1-y_s^2)+(x_s^2+y_s^2-1)^2}+\frac{6\sqrt{6}\,x_s(1+x_s^2-y_s^2-\sqrt{2/3}\,\lambda\,x_sy_s^2)}{4(1-y_s^2)+(x_s^2+y_s^2-1)^2}.\nonumber\eea

The structure of the dynamical system \eqref{asode-vac} entails that the semi-infinite regions $\Psi^+=\{(x_s,y_s,\lambda):x_s>0,\,y_s\geq 0,\,0<\lambda<\infty\}$ and $\Psi^-=\{(x_s,y_s,\lambda):\,x_s<0,\,y_s\geq 0,\,0<\lambda<\infty\}$ are invariant subspaces. This means, that if one gives initial conditions in one of these subspaces, the corresponding orbits of (\ref{asode-vac}) will entirely lay in that subspace. The axes $x_s=0$ and $y_s=0$ are also invariant subspaces. As seen in FIG. \ref{fig-1}, the vertical line $x_s=0$ ($y_s\geq 0$) is a separatrix in the phase space. Hence, the orbits originated from initial conditions in the region $\Psi^-$ will lay entirely in this region. The same is true for orbits in the region $\Psi^+$.

%%%%%%%%%%%%%%%%%%%%%%%%%%%%%%%%%%%%%%%%%%%%%%%%%%%%%

\subsection{New variables and phase space structure}
  
Another interesting thing one may read off from FIG. \ref{fig-1} is that, depending on the initial conditions, the phase space orbits may originate either at the infinities $x_s\rightarrow\pm\infty$, or at the big bang $(x_s,y_s,\lambda)=(0,0,\lambda)$ $\Rightarrow\;z=0$ (see Eq. (\ref{fr-g0-exp})). This means that the variables $x_s$, $y_s$ are unbounded, which poses a problem for the standard variables (\ref{xy-var}), since one or several critical points at infinity may be lost. A way out is to seek for new bounded variables so that all of the possible equilibrium points are ``visible''. Given that in the bottom line of  FIG. \ref{fig-1}, the vertical line $x_s=0$, is a separatrix, one may investigate the dynamics in the invariant subspaces $\Psi^-$ and $\Psi^+$, separately. Accordingly, one may introduce the new bounded variables defined in Eq. (\ref{n-var}). The corresponding phase space where to look for equilibrium points: $\Phi_\text{whole}=\Phi^-\cup\Phi^+$, is the union of the following bounded planes:
 
\bea &&\Phi^+=\{(x_+,y,v):0\leq x_+\leq 1,0\leq y\leq 1,0 \leq v \leq 1\},\nonumber\\
&&\Phi^-=\{(x_-,y,v):-1\leq x_-\leq 0,0\leq y\leq 1,0\leq v\leq 1\}.\nonumber\eea The following separatrices can be identified: 

\bea &&\text{sep}^0:=\left(0,y,v \right),\nonumber\\
&&\text{sep}^+:=\left(x_+,\frac{x_+}{x_++\sqrt{2x_+-1}},v\right),\nonumber\\
&&\text{sep}^-:=\left(x_-,\frac{x_-}{x_--\sqrt{-2x_--1}},v\right).\label{separatrix}\eea Notice that on the separatrices sep$^0$ and sep$^\pm$, $z=1$, i. e., $\sigma_0 H^2=0$, so that either we deal with the static universe there ($H=0$), or, if $\sigma_0=0$, the standard quintessence model with exponential potential -- basically exponential quintessence \cite{wands} -- is recovered. Moreover, for $z=0$ $\Leftrightarrow$ $\sigma_0H^2\rightarrow\infty$, which means that either there is a cosmological singularity there ($H\rightarrow\infty$), or the cubic derivative interaction is decoupled from the gravitational interactions ($\sigma_0\rightarrow\infty$).

%%%%%%%%%%%%%%%%%%%%%%%%%%%%%%%%%%%%%%%%%%

\begin{table*}\centering
\begin{tabular}{||c||c|c|c|c|c|c|}
\hline\hline
Crit. Point        & $x_\pm$   &   $y$   &    $v$     & Existence     & $\omega_\text{eff}$   & $q$  \\
\hline\hline 
$P_{1v}^{\pm}$     & $\pm 1$   &    1    &    $v$      & always     &     1/5    &     4/5    \\
\hline
$P_{2v}^{\pm}$     & $\pm 1$   &    0    &    $v$      & always     &     5    &     8    \\
\hline
$P_{3v}^{\pm}$     & $\pm 1$   & $1/2$   &    $v$  & always     &     $-1$    &    $-1$     \\
\hline 
$P_{4v}^{\pm}$     & $\pm 1/2$   & $1$   &    $ \frac{1}{\lambda_* + 1}$       & always     &     1    &     2     \\
\hline
$P_{5v}^{\pm}$     & $\frac{\pm \sqrt{6}}{\sqrt{6} \pm \lambda_*}$  & $\frac{\sqrt{6}}{\sqrt{6}+\sqrt{6-\lambda_*^2}}$   &   $\frac{1}{\lambda_* + 1}$        & $\lambda_*^2 \leq 6$      &   $-1+\lambda_*^2/3$  &    $-1+\lambda_*/2$     \\
\hline
$P_{6v}^{\pm}$  & $\frac{\pm\lambda_*}{\lambda_*\mp 2\sqrt{6}}$ & 0 & $\frac{1}{\lambda_*+1}$ & if $x_\pm\lambda_*<0$, i. e., if $0\leq x_+\leq 1$, $\lambda_*<0$ &-3&-4 \\
                &   &   &   &\;\;\;\;\;\;\;\;\;\;\;\;\;\;\;\;\;\;\;\;\;\;or if $-1\leq x_-\leq 0$, $\lambda_*>0$ & &  \\
\hline
$P_{7v}$     & $0$   & $0$   &    $1$  & $\lambda_*=0$     &     -3    &    -4     \\
\hline
$P_{8v}$     & $0$   & $1$   &    $1$       & $\lambda_*=0$   &     -3    &     -4     \\
\hline\hline
\end{tabular}\caption{Cubic galileon vacuum for constant coupling $\sigma=\sigma_0$. The physically meaningful critical points of the autonomous system \eqref{ode-vac}, \eqref{def-s-vac}, together with the existence conditions, the effective EOS $\omega_\text{eff}$ and the deceleration parameter $q$ are shown.}\label{tab-3}\end{table*}

\begin{table*}\centering
\begin{tabular}{||c||c|c|c|c|}
\hline\hline
Crit. Point   & $\lambda_1$   &   $\lambda_2$  &    $\lambda_3$     & Stability    \\
\hline\hline 
$P_{1v}^{\pm}$  &      9/5  &   $6/5$   &    0   &   unstable node      \\ 

\hline
$P_{2v}^{\pm}$  &      12    &    -9    &   0      & saddle        \\

\hline
$P_{3v}^{\pm}$  &      -9/2   &    -3/2   &    0   &    stable node      \\
                &               &           &       &  (numerical investigation)  \\
\hline
$P_{4v}^{\pm}$  &      -6   &  $\mp \sqrt{6} \lambda_*$  &   $3\mp\sqrt\frac{3}{2}\lambda_*$  &  stable node if $\lambda_*^2>6$  \\
                &           &                            &                                    &  saddle otherwise  \\
\hline
$P_{5v}^{\pm}$  &    $-\lambda_*^3\Gamma_*$   &  $-\lambda_*^2$   &   $-3+\lambda_*^2/2$  &  stable node if $\lambda_*\Gamma_*>0$ \\
                &                             &                   &                       &  saddle otherwise  \\
\hline
$P_{6v}^{\pm}$  &      -12   &   -3    &  $-12 \lambda_*\Gamma_*$   & stable node if $\lambda_*\Gamma_*>0$   \\
                &            &         &                            &  saddle otherwise  \\
\hline 
$P_{7v}$  &     $6$   &    $3$    &    undefined       & unstable node  \\
\hline
$P_{8v}$  &     $-6$   &    $3$    &    undefined       & saddle  \\
\hline\hline
\end{tabular}\caption{Eigenvalues of the linearization matrix and the corresponding stability of the critical points of the dynamical system \eqref{ode-vac}, \eqref{def-s-vac}.}\label{tab-4}\end{table*}

%%%%%%%%%%%%%%%%%%%%%%%%%%%%%%%%%%%%%%%%

\subsection{Dynamical systems analysis}

In terms of the set of variables (\ref{n-var}) the dynamical system for the galileon vacuum with constant coupling $\sigma=\sigma_0$ is given by 

\bea &&x'_\pm=-\frac{x_\pm^2}{\sqrt 6}\left(\frac{\ddot\phi}{H^2}\right)_\pm+x_\pm(1\mp x_\pm)\left(\frac{\dot H}{H^2}\right)_\pm,\nonumber\\
&&y'=y(1-y)\left[\sqrt\frac{3}{2}\left(\frac{1-v}{v}\right)\left(\frac{1\mp x_\pm}{x_\pm}\right)+\left(\frac{\dot H}{H^2}\right)_\pm\right],\nonumber\\
&&v'=\sqrt{6}\left(\frac{1\mp x_\pm}{x_\pm}\right)\,F(v),\label{ode-vac}\eea where as before $F(v)\equiv v^2f(v)$, and:

\bea &&\left(\frac{\dot H}{H^2}\right)_\pm=\frac{3\left[3x_\pm^4(1-y)+4(1\mp x_\pm)^2x_\pm^2y^3-10x_\pm^2+2x_\pm-2)x_\pm^2y^2-(1\mp x_\pm)^4y^4\right]}{-4x_\pm^4y+x_\pm^4+4(x_\pm(2\mp x_\pm)-1)x_\pm^2y^3-2(1\mp x_\pm)^2x_\pm^2y^2-(1\mp x_\pm)^4y^4},\nonumber\\
&&\left(\frac{\ddot\phi}{H^2}\right)_\pm=\frac{-3\sqrt{6}(1\mp x_\pm)\left[x_\pm^4(y^2+2y-1)(3y^2-2y+1)-8x_\pm^2y^4(1\mp x_\pm)-4y^4(1\mp x_\pm)\right]}{-4x_\pm^4y+x_\pm^4+4(x_\pm(2\mp x_\pm)-1)x_\pm^2y^3-2(1\mp x_\pm)^2x_\pm^2y^2-(1\mp x_\pm)^4y^4}.\label{def-s-vac}\eea Recall that, due to the constraint \eqref{fr-g0-exp}, this dynamical system is a 3-dimensional one, unlike the one for the case with matter that is a 4-dimensional dynamical system.

The real-valued, physically meaningful critical points $(x_{\pm *},y_*,v_*)$ of the dynamical system \eqref{ode-vac}, \eqref{def-s-vac}, together with their existence conditions, the value of the effective EOS: $\omega_\text{eff}$, and the deceleration parameter $q$, can be found in TAB. \ref{tab-3}. In TAB. \ref{tab-4} we summarize the eigenvalues of the linearization matrices and the stability properties of these critical points. As in the previous section, the analysis for the non hyperbolic critical points is performed by means of the numerical investigation.

\begin{enumerate}

\item The big bang solution: $P_{1v}^\pm:(\pm 1, 1,v)$. Since in this case ($\sigma_0\neq 0$): $$z=\frac{1}{H^2\sigma_0+1}=0\;\Rightarrow\;H\rightarrow\infty,$$ this equilibrium configuration should be associated with the initial big bang. In terms of the standard (unbounded) normalized variables ($x_s$,$y_s$,$\lambda$), the equilibrium manifold $P_{1v}^\pm\Rightarrow (0,0,\lambda)$, and, according to TAB. \ref{tab-3} its existence is independent of the specific form of the potential. Given that, until a specific potential is picked up, this is an equilibrium (linear) manifold, the eigenvalue of the linerization matrix that is aligned with the $v$-direction ($\lambda_3$ in TAB. \ref{tab-4}) vanishes. The same holds true for the critical manifolds $P_{2v}^\pm$ and $P_{3v}^\pm$. We want to underline that, once a specific potential is chosen, given that then $v$ (or, equivalently, $\lambda$) takes a specific value, the corresponding equilibrium configuration is an isolated point in $P_{1v}^\pm$. For this equilibrium manifold, since $q=4/5$, then: $$\frac{\dot H}{H^2}=-\frac{9}{5}\;\Rightarrow\;H=\frac{5/9}{t-t_0}\;\Rightarrow\;a(t)\propto(t-t_0)^{5/9},$$ i. e., as already said, $P_{1v}^\pm$ is to be associated with a pure galileon big bang singularity at some initial time $t_0$ (compare with the point $P^\pm_1$ in TAB. \ref{tab-1} which is associated with a matter-dominated big bang instead). This unstable solution which corresponds to a saddle point in the phase space has not analogues in the standard quintessence model.

\item The super-decelerated expansion solution: $P^{\pm}_{2v}$ corresponds to a transient stage of the cosmological evolution. Since $q=8$, $$\frac{\dot H}{H^2}=-9\;\Rightarrow\;a(t)\propto (t-t_0)^{1/9}.$$ The existence of this equilibrium point is independent of the specific form of the potential and it is always a saddle.

\item The de Sitter solution: $$P^{\pm}_{3v}:(\pm 1,1/2,v)\Rightarrow \dot\phi=0,\;H=\sqrt{V_0/3},$$ corresponds to a local stable solution of the dynamical system \eqref{ode-vac}, so that it can be the late-time attractor for a non-empty set of phase space orbits. In this case the parameter $z$ is undefined since, as long as the de Sitter equilibrium configuration is approached along the separatrices sep$^\pm$, then $z=1$. Meanwhile, for other approaching directions $z=0$. The de Sitter solution does not arise in standard exponential quintessence, unless $\lambda=0$ (constant potential case), so that its existence for any $\lambda\neq 0$ is a genuine consequence of the galileon coupling $\sigma_0\neq 0$. As previously found in \cite{rtgui}, for the cubic galileon vacuum there is no self-accelerated solution.

\item The stiff matter (decelerated expansion) solution: $$P^{\pm}_{4v}:\left(\pm\frac{1}{2},1,\frac{1}{\lambda_*+1}\right)\;\Rightarrow\;H^2=\dot\phi^2/6,$$ for which $z=1$, is also found in the more general scenario when the matter component is present. As it was for the quintessence, this unstable equilibrium configuration is not relevant at late times.

\item The critical points that are dominated by the scalar field $$P^\pm_{5v}:\left(\frac{\pm\sqrt{6}}{\sqrt{6}\pm\lambda_*},\frac{\sqrt{6}}{\sqrt{6}+\sqrt{6-\lambda_*^2}},\frac{1}{\lambda_*+1}\right),$$ have the same properties as in the exponential quintessence model \cite{wands}. These correspond to scaling of the kinetic and potential energies of the scalar field: $$\frac{\dot\phi^2}{2V}=\frac{\lambda_*^2}{6-\lambda_*^2}.$$ Whenever the bound $\lambda_*\Gamma_*<0,$ is met, the equilibrium states $P^\pm_{5v}$ are stable, and so, these are important at late times. Otherwise, these are saddle critical points, representing transitory states.

\item The phantom solution $$P_{6v}^\pm:\left(\frac{\pm\lambda_*}{\lambda_* \mp 2\sqrt{6}},0,\frac{1}{\lambda_*+1}\right),\;z=0,$$ is a stable critical point (a local attractor) whenever $\lambda_*\Gamma_*>0$ and it is a saddle point otherwise. This equilibrium state exists if $x_\pm\lambda_*<0$, i. e., if: $$0\leq x_+\leq 1\;\text{and}\;\lambda_*<0,$$ or if $$-1\leq x_-\leq 0\;\text{and}\;\lambda_*>0,$$ i. e., for monotonically growing potentials. In order to illustrate the phantom behavior of this solution let us to choose the patch where $0\leq x_+\leq 1$. In this case the solution exists only for negative $\lambda_*<0$. Let us set $\lambda_*=-\kappa$, with $\kappa>0$. We have that 

\bea &&x_+=\frac{\kappa}{\kappa+2\sqrt{6}}\;\Rightarrow\;\dot\phi=\frac{12}{\kappa}\,H,\nonumber\\
&&y=0\;\Rightarrow\;\frac{\sqrt{V}}{\sqrt{3}H}\rightarrow\infty,\nonumber\\
&&z=0\;\Rightarrow\;\sigma_0H^2\rightarrow\infty.\nonumber\eea From the first equation above it follows that $$\phi(a)=\frac{12}{\kappa}\,\ln a+\phi_0,$$ where $\phi_0$ is an arbitrary integration constant. Additionally, since for this critical point $q=-4$ (it is a super-accelerated solution), then: $\dot H=3H^2$, so that $$H(t)=\frac{1}{3(t_f-t)}\;\Rightarrow\;a(t)=\frac{a_0}{(t_f-t)^{1/3}},$$ where $-3t_f$ and $\ln a_0$ are arbitrary integration constants, and $t\leq t_f$. Besides, for the effective energy density we have that: $$\rho_\text{eff}(t)=3H^2(t)=\dot H(t)=\frac{1}{3(t_f-t)^2},$$ where the phantom behavior is evident from the fact that the energy density of the cubic galileon unboundedly grows up with $t$. Given that $a(t)$, $H(t)$, $\dot H(t)$, and $\rho_\text{eff}(t)$, all blow up at $t=t_f$, i. e., in a finite time into the future, a big rip singularity \cite{odintsov, ruth} may be the inevitable fate of the cosmic evolution if the bound $\lambda_*\Gamma_*>0$ is met. For definiteness let us choose the exponential potential (for the combination of exponentials the analysis is similar): $V\propto\exp(\kappa\phi)\propto a^{12}\propto(t_f-t)^{-4}$. It is verified that, as $t\rightarrow t_f$ asymptotically, $$H^2\sigma_0\propto(t_f-t)^{-2}\rightarrow\infty\Leftrightarrow\;z=0,$$ so that $$\frac{\sqrt V}{H}\propto(t_f-t)^{-1}\rightarrow\infty\Leftrightarrow\;y=0,$$ as required. Recall that at this point $y=0$ $\Rightarrow\;\sqrt{V}/H\rightarrow\infty$. Another way to explain the arising of the latter limit -- without specifying the functional form of the potential -- is by noticing that, since $\dot\phi=12H/\kappa$, the Friedmann equation can be written as: $$V=\left(3-\frac{\alpha^2}{2}\right)H^2+3\alpha^3\sigma_0H^4,$$ where we have set $\alpha\equiv 12/\kappa$. As seen, the self-interaction galileon potential $V(\phi)$ asymptotically approaches to $$V\propto H^4\;\Rightarrow\;\frac{\sqrt{V}}{H}\propto H\rightarrow\infty,$$ as required by the consistency of the phantom solution.

\item The equilibrium point $P_{7v}:(0,0,1)$ is associated with a super-accelerated contraction of the universe and corresponds to a unstable node in the phase-space. This super-accelerated solution exists only for constant potentials or for potentials that approach asymptotically to a constant: $\lambda_*=0$. From equation \eqref{fr-g0-exp} it can be seen that, for the present case $z=1$, and this is achieved only if $H^2\sigma_0=0$. This means that asymptotically $H\rightarrow 0$, which is consistent with $x_\pm=0$ ($x_s\rightarrow\infty$) if $\dot\phi\rightarrow 0$ more slowly than $H$, and it is also consistent with $y=0$ ($y_s\rightarrow\infty$). For this case, since $\dot H=3H^2$, we have that $$H(t)=-\frac{1/3}{t+C_0},$$ where $C_0$ is an integration constant. As seen the above choice of the solution $H=H(t)$ holds the required asymptotics: $H(t)\rightarrow 0$ as $t\rightarrow\infty$. The ''$-$'' sign entails that the solution depicts a contracting universe. Since in the present paper we are interested in expanding solutions, this one (and the one below) that belong in the boundary of the phase space, may be safely ignored. In spite of this we have included these solutions in order to illustrate the complexity of the cubic galileon vacuum, since none of the solutions $P^\pm_{6v}$, $P_{7v}$ and $P_{8v}$ are found if the galileon vacuum is filled with standard matter degrees of freedom.

\item The solution $P_{8v}:(0,1,1)$ is very similar to the former one. This represents a super-accelerating contracting phase of the cosmic evolution as well. This solution -- the same is true for the above equilibrium state $P_{7v}$ -- has no impact in the late-time dynamics. As before the asymptotics $H\rightarrow 0$ is required for consistency of the solution. The only difference of this critical point with $P_{7v}$ is that, while in the latter equilibrium state the super-accelerated contraction is fueled by the galileon with (asymptotically) constant potential, in the present case, since $y=1$ $\Rightarrow\;y_s=0$, the contraction is driven by the pure kinetic energy of the galileon (vanishing potential). 

\end{enumerate} 

The phantom solution above: $P^\pm_{6v}$, is perhaps the most distinctive feature of the complexity of the cubic galileon vacuum.\footnote{ The super-accelerated solutions $P_{7v}$ and $P_{8v}$ are not of importance for our analysis since these correspond to contracting universe.} This may have implications for the late-time asymptotics and, hence, may be of importance for the future destiny of our universe. The fact that the addition of standard matter degrees of freedom, say dust-like dark matter, screens this vacuum effect is a very interesting example of the physical role of the cubic self-interaction of the galileon $\propto(\nabla^2\phi)(\nabla\phi)^2$, that is intimately linked with the cosmological Vainshtein screening mechanism \cite{vainshtein, khoury-rev}.

%%%%%%%%%%%%%%%%%%%%%%%%%%%%%%%%%%%%%%%%

\section{Discussion}\label{sec6}

In order to illustrate which is the effect of the cubic interaction of the galileon on the vacuum, let us to briefly expose the results of the dynamical systems analysis of quintessence models with potentials beyond the exponential one \cite{Fang}. For simplicity, in addition to the quintessence, we assume pressureless (dust-like) matter field. In terms of the standard normalized variables $x_s$, $y_s$ and $\lambda$, the dynamical system for this case reads:

\bea &&x'_s=-3x_s+\frac{3}{2}\,x_s\left(1+x_s^2-y_s^2\right)+\sqrt\frac{3}{2}\,\lambda y_s^2,\nonumber\\
&&y'_s=-\sqrt\frac{3}{2}\,\lambda x_sy_s+\frac{3}{2}\,y_s\left(1+x_s^2-y_s^2\right),\nonumber\\
&&\lambda'=-\sqrt{6}\,x_s\lambda^2f(\lambda).\label{wands-asode}\eea The Friedmann constraint is written in the following form:

\bea \Omega_m=1-x_s^2-y_s^2,\;0\leq\Omega_m\leq 1.\label{wands-friedmann-const}\eea The physically meaningful phase space is the infinite cylinder

\bea &&\Psi_Q:=\{(x_s,y_s,\lambda):|x_s|\leq 1,\,y_s\geq 0,x_s^2+y_s^2\leq 1,-\infty<\lambda<\infty\}.\label{ph-sp-q}\eea 

The three-dimensional dynamical system \eqref{wands-asode} reduces to a two-dimensional system when $f(\lambda)=0,$ (as it is for the exponential potential case) and has been properly studied in \cite{wands}. For potentials beyond the exponential the corresponding study has been published in \cite{Fang}. In this case all of the physically meaningful equilibrium points $P^*:(x^*_s, y^*_s,\lambda^*)$ of (\ref{wands-asode}) are located within the cylinder $\Psi_Q$ in \eqref{ph-sp-q}. Since the phase space is unbounded along the $\lambda$-direction, it might exist one (or more) critical points at infinity, and thus it would be necessary to perform an additional analysis -- like the Poincar\'e projection procedure -- to uncover the asymptotic structure. Fortunately this is not the case and all of the real-valued, physically meaningful critical points fit into a finite part of the above cylinder. These equilibrium points together with their existence conditions can be found in TAB. \ref{tab-fang}, while in TAB. \ref{tab-fang-stab} we summarize the eigenvalues of the corresponding linearization matrices and the stability properties of the critical points.

%%%%%%%%%%%%%%%%%%%%%%%%%%%%%%%%%%

\begin{table*}[tbh]\centering
\begin{tabular}{||c||c|c|c|c|c|c|}
\hline\hline
Crit. Point        &  Existence   &   $\Omega_m$   &   $\omega_\text{eff}$   &   $q $ \\
\hline\hline
$P_1^*: (0,0,\lambda)$   &   always     &       1        &        und.         &   1/2  \\
\hline
$P_2^*: (\pm 1,0,\lambda_*)$     &   always     &       0        &      1      &   2  \\
\hline
$P_3^*: (\frac{\sqrt{3}}{\sqrt{2} \lambda_*},\frac{\sqrt{3}}{\sqrt{2} \lambda_*},\lambda_*)$     &   $\lambda_*^2 \geq 3$     &       $1-\frac{3}{\lambda_*^2}$       &        0        &   2  \\
\hline
$P_4^*: (\frac{\lambda_*}{\sqrt{6}},\sqrt{1-\frac{\lambda_*^2}{6}},\lambda_*)$     &   $\lambda_*^2 <6$     &    0      &        $-1+\frac{3}{\lambda_*^2}$         &   $-1+\frac{3}{\lambda_*^2}$ \\
\hline
$P_5^*: (0,1,0)$         &   always     &       0        &        -1         &   -1  \\
\hline\hline
\end{tabular}\caption{Critical points $P^*:(x_s^*,y_s^*,\lambda^*)$ of the dynamical system \eqref{wands-asode} together with their existence and physical properties. Here $\lambda_*$ corresponds to any node of the function in the RHS of \eqref{ode-l}.}\label{tab-fang}\end{table*}

%%%%%%%%%%%%%%%%%%%%%%%%%%%%%%%%%%%%%%%%%%%%%%%%%%%%%%%%%%%%%%%%%%%%%

\begin{table*}[tbh]\centering
\begin{tabular}{||c||c|c|c|c|c|c||}
\hline\hline
Crit. Point      &  $\lambda_1$   &  $\lambda_2$   &   $\lambda_3$   &   Stability \\
\hline\hline
$P_1^*$   &   -3/2   &     3/2      &       0        &   saddle  point\\
\hline
$P_2^*$  & 3 &  $3\mp\sqrt\frac{3}{2}\lambda_*$ & $\mp\sqrt{6}\lambda_*^2\Gamma_*$ &  unstable if $\pm\lambda_*>-\sqrt{6}$ and $\Gamma_*>0$\\
&&&& saddle otherwise \\
\hline
$P_3^*$   &   -3 & $-\frac{3}{4}+\alpha$  &   $- \frac{3}{4}-\alpha$  &  stable if $\lambda_*\Gamma_* >0 $  \\
&&&& saddle otherwise \\
\hline
$P_4^*$   & $-\lambda_*^3\Gamma_*$ & $-3+\lambda_*^2$  &  $-3+\frac{\lambda_*^2}{3}$ & stable if $\lambda_*^2<3$ and $\lambda_*\Gamma_*>0$ \\
\hline
$P_5^*$   &   -3     & $-\frac{3}{2}+\beta$ & $-\frac{3}{2}+\beta$ & stable if $f(0)>0$  \\
&&&& saddle if $f(0)<0$ \\
\hline\hline\end{tabular}\caption{Stability of the critical points of the dynamical system \eqref{wands-asode}. Here we have defined $\alpha\equiv\frac{3}{4\lambda_*}\sqrt{24-7\lambda_*^2}$ and  $\beta\equiv\frac{\sqrt{3}}{2}\sqrt{3-4f(0)}$, meanwhile $f(0)$ denotes the value of the function $f(\lambda)$ evaluated at $\lambda=0$.}\label{tab-fang-stab}\end{table*}

%%%%%%%%%%%%%%%%%%%%%%%%%%%%%%%%%%

There are two equilibrium points associated with the presence of dark matter: $P_1^*$, which is related to the matter dominated solution and  the scalar field-matter scaling solution $P_3^*$. The remaining singular points: the stiff matter point $P_2^*$, the scalar field dominated solution $P_4^*$ and the de Sitter universe $P_5^*$, are found even in the absence of a matter component, i. e., these correspond to the quintessential vacuum. In other words, the phase space asymptotic structure of the quintessence vacuum is characterized by the equilibrium points $P_2^*$, $P_4^*$ and $P_5^*$ exclusively. This is corroborated by setting $\Omega_m=0$ in \eqref{wands-friedmann-const}: $$\Omega_m = 0\;\Rightarrow\;x_s^2+y_s^2 = 1.$$ This relationship between the variables $x_s$ and $y_s$ leads to a reduction of the dimensionality of the dynamical system from 3 to 2. In other words, one is left with a coupled autonomous ODE:

\bea &&x'_s=\left(\sqrt\frac{3}{2}\,\lambda-3x_s\right)\left(1-x_s^2\right),\nonumber\\
&&\lambda'=\sqrt{6}\,x_s\lambda^2f(\lambda).\label{quint-vac-asode}\eea The only critical points $P_*:(x_s^*,\lambda_*)$ of this dynamical system are: $P_2^*:(\pm 1,0)$ with $y_s^*=0$, $P_4^*:(0,0)$ with $y_s^*=1$ and $$P_5^*:\left(\frac{\lambda_*}{\sqrt 6},\lambda_*\right)\Rightarrow y_s^*=\sqrt{1-\frac{\lambda_*^2}{6}},$$ respectively. In consequence, of the five equilibrium points of the dynamical system \eqref{wands-asode} corresponding to the quintessence with background matter (see TAB. \ref{tab-fang}), only three critical points: $P_2^*$, $P_4^*$ and $P_5^*$, survive in the particular vacuum case. What we have just demonstrated is that, if take the continuous limit $\Omega_m\rightarrow 0$ of the dynamical system \eqref{wands-asode} with its corresponding phase space asymptotics, we obtain the dynamical system \eqref{quint-vac-asode} together with its corresponding (whole) phase space structure.

\subsection{The cosmological vDVZ discontinuity}

The above result is to be expected in a scalar tensor theory without derivative interactions but not in a theory like \eqref{action}. If compare the results for standard quintessence in TAB. \ref{tab-fang} with those for cubic quintessence in TAB. \ref{tab-1}, we see that but for the big-bang solution $P_1^\pm$ in TAB. \ref{tab-1}, the results coincide so that for the present model the late-time dynamics is essentially the same as for standard quintessence. This is to be contrasted with our finding that, thanks to the cubic self-interaction of the galileon: $\sigma_0\dot\phi^2(\ddot\phi+3H\dot\phi)$, the asymptotic dynamics of vacuum is richer (more critical points) than the one in the presence of background matter. This is easily seen from TAB. \ref{tab-3} for the cubic galileon vacuum, where the critical points $P^\pm_{6v}$ -- together with the uninteresting equilibrium points $P_{7v}$ and $P_{8v}$ -- have no analogues in the TAB. \ref{tab-1} corresponding to the cubic galileon with background matter.

This latter result -- first reported in \cite{rtgui} for the exponential potential case -- can be related with a kind of cosmological vDVZ discontinuity. Actually, if we remove from TAB. \ref{tab-1} the equilibrium points that are associated with the background matter: the matter-dominated big bang (point $P^\pm_1$), the dark matter domination solution ($P^\pm_2$) and the matter-scaling solution ($P^\pm_6$), we are left with the only galileon vacuum solutions of the dynamical system \eqref{ode-mat}: the de Sitter solutions (critical point $P^\pm_3$), the stiff-matter ($P^\pm_4$) and the quintessence dominated solutions ($P^\pm_5$). But the cubic galileon vacuum dynamical system \eqref{ode-vac} is a particular case of \eqref{ode-mat} when we set $\Omega_m=0$, so that the above mentioned vacuum solutions (points $P^\pm_3$, $P^\pm_4$ and $P^\pm_5$), should be the only physically meaningful equilibrium points of \eqref{ode-vac}, which is not the case as shown in TAB. \ref{tab-3}. In other words, we can not get the whole phase dynamics of the cubic galileon vacuum in the continous limit $\Omega_m\rightarrow 0$ of the more general dynamical system \eqref{ode-mat} corresponding to the cubic galileon with background matter (TAB. \ref{tab-1}). This is what we call as the cosmological version of the vDVZ discontinuity.

The vDVZ can be avoided if assume that the cubic self-interactions of the galileon are somehow screened by its interactions with the background matter so that, for instance, the phantom solution that may affect the late time cosmic dynamics of the galileon vacuum is erased from the phase space. In consequence, in the presence of matter degrees of freedom (in addition to the galileon) the late-time dynamics of the model is essentially the same as for standard quintessence \cite{genly}.

\subsection{The cosmological Vainshtein screening}

The above cosmological screening effect is similar to the cosmological versions of the Vainshtein mechanism explained in \cite{chow} (see also \cite{silva-kazuya}) that operates at high energies when the non-linear terms in the equations of motion dominate over the linear one, thus leading to the recovery of general relativity. In the case of the phantom vacuum solution $P^\pm_{6v}$, since it is related with a big rip singularity where $$\dot H\sim H^2\sim a^6\sim\rho_\text{eff}\rightarrow\infty,$$ what happens is that, at late times, the universe enters a high energy regime where the cubic term dominates. This results in that the galileon decouples from the other matter degrees of freedom, so that we are left effectively with general relativity \cite{chow}. 

For the super-accelerated contracting solutions $P_{7v}$ and $P_{8v}$, the explanation of the screening effect is a bit different since in this case the effective energy density (the cubic galileon's energy density) dilutes with the contraction while the matter energy density grows up with the cosmic time: $\rho_m\sim a^{-3}\propto t$. In order to expose our reasoning line in this case, let us rewrite the Friedmann equation \eqref{fried} in the following convenient way: 

\bea H^2+\sigma_0\dot\phi^3H=\frac{1}{3}\left(\rho_m+\rho_\phi\right),\label{dgp-fried}\eea where, as before, $\rho_\phi=\dot\phi^2/2+V$. Written in this form the Friedmann equation for the cubic galileon resembles the one for the DGP braneworld \cite{dgp, roy-rev}: $$H^2\pm\frac{1}{r_c}H=\frac{1}{3}\rho_m,$$ where the crossover scale $r_c=G_{(5)}/2G_{(4)}$ is half of the ratio between the 5D and 4D gravitational couplings. If compare this latter equation with \eqref{dgp-fried} one can identify the cubic term $\sigma_0\dot\phi^3$ with the inverse of certain ''crossover'' scale: $r_*=(\sigma_0\dot\phi^3)^{-1}$. In correspondence one may also identify a Vainshtein radius: $r_V=(r_g r_*^2)$, within which the non-linear cubic interaction becomes important. In this latter relationship $r_g$ is the Schwarzschild radius of the universe that is roughly the Hubble scale $r_g\sim H^{-1}$. An acceptable estimate for $r_*$ would be that $r_*\sim H^{-1}$ also. Hence, since for the super-accelerated solutions $H=-1/3(t+C_0)$ -- see the former section -- where $C_0$ is an integration constant which, for simplicity, may be set to zero, then the Vainshtein radius grows up with the cosmic time $$r_V\sim |H|^{-1}\propto t,$$ while the physical distances go like: $d_\text{phys}=ra(t)\propto t^{-1/3}$. As the contraction proceeds, eventually, there will be a regime where the physical distances start becoming smaller than the Vainshtein radius: $d_\text{phys}\lesssim r_V$, so that the non-linear (cubic) self-interaction of the galileon becomes dominating. This leads to the decoupling of the galileon interactions which results in the recovering of general relativity. Although in this demonstration we have assumed the estimate $r_*\sim H^{-1}$, we can see that even without the assumption of any estimate, since in general for $\dot\phi>0$ the crossover scale $r_*=(\sigma_0\dot\phi^3)^{-1}$ decays with the cosmic time, the above conclusion is always true.

\subsection{Speed of gravitational waves}

Before we conclude this section we want to make a comment on one important aspect that has gained interest recently. It is related with the tight constraint on the difference in speed of photons and gravitons 

\bea |c^2_T-c^2|\leq 6\times 10^{-15}c^2,\label{const-gw}\eea where $c_T$ is the speed of the gravitational waves (recall that in this paper $c^2=1$), implied by the announced detection of gravitational waves from the neutron star-neutron star merger GW170817 and the simultaneous measurement of the gamma-ray burst GRB170817A \cite{ligo}. Take for instance, the Horndeski-type theory with kinetic coupling of the scalar field to the Einstein's tensor: $\alpha G_{\mu\nu}\der^\mu\phi\der^\nu\phi$ \cite{gao, sushkov, saridakis-sushkov, matsumoto, granda, germani-prl}. In this theory the squared speed of sound of the gravitational waves is given by \cite{germani}: 

\bea c^2_T=\frac{2+\alpha\dot\phi^2}{2-\alpha\dot\phi^2},\label{c2t-germani}\eea so that, depending on the kinetic energy of the scalar field, either a Laplacian instability develops ($\alpha\dot\phi^2>2$) or the gravitational waves may travel at superluminal speed ($\alpha\dot\phi^2<2$). The fact is that in the mentioned theory the speed of propagation of the gravitational waves may substantially differ from the local speed of light thus rendering the resulting cosmological model incompatible with the above constraint \eqref{const-gw}. In contrast, the present model based on \eqref{action} is not constrained by the above mentioned combined detection of gravitational waves from the neutron star merger GW170817 and the simultaneous measurement of GRB170817A reported in \cite{ligo}, since for the cubic galileon model the speed of propagation of the tensor gravitational waves perturbations exactly coincides with the local speed of light: $c_T=1$.

\subsection{Validity of the assumption $H\geq 0$} 

In this paper we have assumed non-negative $\sigma_0$ and ever expanding universes: $H>0$. As seen from \eqref{h-eq}, for the present cubic galileon model -- according to the Friedmann equation \eqref{fried} -- for the Hubble rate one gets:

\bea H_\pm=-\frac{\sigma_0\dot\phi^3}{2}\pm\sqrt{\frac{\sigma_0^2\dot\phi^6}{4}+\frac{1}{3}\left(\rho_m+\rho_\phi\right)},\nonumber\eea where $\rho_\phi:=\dot\phi^2/2+V$. Since we focused our study in ever expanding universes, we have investigated the '$+$' branch of
the above equation exclusively. A question may naturally arise: what if bouncing solutions are found? In such a case even initially expanding solutions; $H(t_0)>0$, may turn into contracting ones at some later time $t>t_0$: $H(t)<0$, or what if the Hubble rate is simply a monotonically decreasing function, i. e., $\dot H<0$, so that at some $t_*$, $H(t_*)=0$ and after that $H<0$, so that our assumption that $H\geq 0$ may not be valid for the entire cosmic history? In other words: what if the value $H=0$ is crossed by given solutions of the present model? In the reference \cite{qiu}, for instance, the authors found bouncing solutions in a conformal galileon model that shares certain resemblance to ours. Below we shall show that in our model the value $H=0$ is not crossed by any solution that is driven by non-negative potentials (the case of interest for cosmological applications). The demonstration will be based in the dynamical systems analysis of the cosmological equations \eqref{fried}-\eqref{ecc} for the cubic galileon. We shall work in some state space with coordinates $\dot\phi$, $H$ and $\lambda$ -- the slope of the self-interaction potential.

Let us start our demonstration with the simplest situation: the cubic galileon vacuum. In this case the motion equations read:

\bea &&3H^2=\frac{X^2}{2}\left(1-6\sigma_0 XH\right)+V,\label{fried-eq'}\\
&&-2\dot H=X^2-3\sigma_0 X^3H+\sigma_0 X^2\dot X,\label{ray-eq'}\\
&&(1-6\sigma_0 XH)\dot X+3XH-9\sigma_0 X^2H^2-3\sigma_0 X^2\dot H=\lambda V,\label{kg-eq'}\eea where we have introduced the variable $X\equiv\dot\phi$ and the slope of the potential\footnote{In general $\lambda$ is a function of $\phi$.} $\lambda:=-V_{,\phi}/V$ is defined in \eqref{lamb}. From \eqref{fried-eq'} we can write:

\bea V=3H^2+3\sigma_0 X^3H-\frac{X^2}{2},\label{v-eq}\eea and substitute $V$ back into \eqref{kg-eq'}. Then we can write the equations \eqref{ray-eq'} and \eqref{kg-eq'} in the form of the following dynamical system in the variables $X,H,\lambda$:

\bea &&\dot X=\frac{-3X(2H+\sigma_0 X^3)(1-3\sigma_0 XH)+\lambda\left(6H^2+6\sigma_0 X^3H-X^2\right)}{2-12\sigma_0 XH+3\sigma_0^2X^4},\nonumber\\
&&\dot H=-\frac{X^2(1-3\sigma_0 XH)(1-9\sigma_0 XH)+\sigma_0 X^2\lambda\left(3H^2+3\sigma_0 X^3H-X^2/2\right)}{2-12\sigma_0 XH+3\sigma_0^2X^4},\label{dyn-syst}\\
&&\dot\lambda=-Xf(\lambda),\nonumber\eea where the function $f$ is given by \eqref{f-eq} while $\Gamma$ is defined in \eqref{Gamma}: $$f(\lambda)\equiv\lambda^2(\Gamma-1),\;\;\Gamma:=VV_{,\phi\phi}/V^2_{,\phi}.$$ The phase space where to look for critical points of the dynamical system \eqref{dyn-syst} is $\Psi=\{(X,H,\lambda)\in\text{R}^3\}$. An additional physical requirement we impose on the phase space is that the potentials that can drive viable cosmological behavior are non-negative: $V\geq 0$, i. e., 

\bea 3H^2+3\sigma_0 X^3H-\frac{X^2}{2}\geq 0\;\Rightarrow\;H_\pm=-\frac{\sigma_0 X^3}{2}\pm\sqrt{\frac{\sigma_0^2 X^6}{4}+\frac{X^2}{6}},\label{cond-1}\eea where the regions with positive $V>0$ are those depicted by the conditions: $H>H_+$ and $H<H_-$, respectively. It is easily checked that the lineal manifold ${\cal M}=(X,H,\lambda)=(0,0,\lambda)$ is a critical manifold of the dynamical system \eqref{dyn-syst}. 

The possibility to cross the value $H=0$ is split into two complementary conditions on the derivative of the galileon: either $X=0$, or $X\neq 0$, respectively. There is no other possibility at all. In the former case: $X\equiv\dot\phi=0$, $H=0$ ($\Rightarrow$ $\dot H=0$), since points in ${\cal M}$ are critical points of the dynamical system \eqref{dyn-syst}, those orbits of the phase space that approach ${\cal M}$ can not cross $H=0$. In other words: orbits that approach to the plane $(X,H,\lambda)=(0,H,\lambda)$ can not cross the value $H=0$ since points with $X=H=0$ are critical points that belong in ${\cal M}$. In the complementary case: $X\neq 0$ ($\dot\phi\neq 0$), the phase space orbits may in principle cross the value $H=0$, including the possibility of a non-singular bounce if $\dot H>0$. However, in this case, from the Friedmann equation \eqref{fried-eq'}, it follows that: $X^2/2+V=0$, i. e., it is necessary that the potential $V=-X^2/2<0$ be a negative quantity. We recall that in our paper we are considering strictly non-negative potentials $V\geq 0$. Hence, the assumption that $H\geq 0$ on which the present investigation is based, is a valid assumption. 

Although the above demonstration is valid for the vacuum galileon with the cubic self-interaction, it is easily generalized to the case with background matter. In this latter case we notice that the only possibility that the crossing of the $H=0$ happens is when $\dot\phi\neq 0$, since $(X,H,\lambda)=(0,0,\lambda)$ is a critical manifold even in the presence of matter degrees of freedom other than the galileon itself. But, from the Friedmann equation: $3H^2=\rho_\text{mat}+X^2(1-6\sigma_0 XH)/2+V$, it follows that at $H=0$ ($X\neq 0$): $V=-\rho_\text{mat}-X^2/2$, the potential should be an strictly negative quantity.

Below we give additional numerical support to the above demonstration by including the numeric investigation of two particular cases.

\begin{figure}\centering
\includegraphics[width=5.5cm]{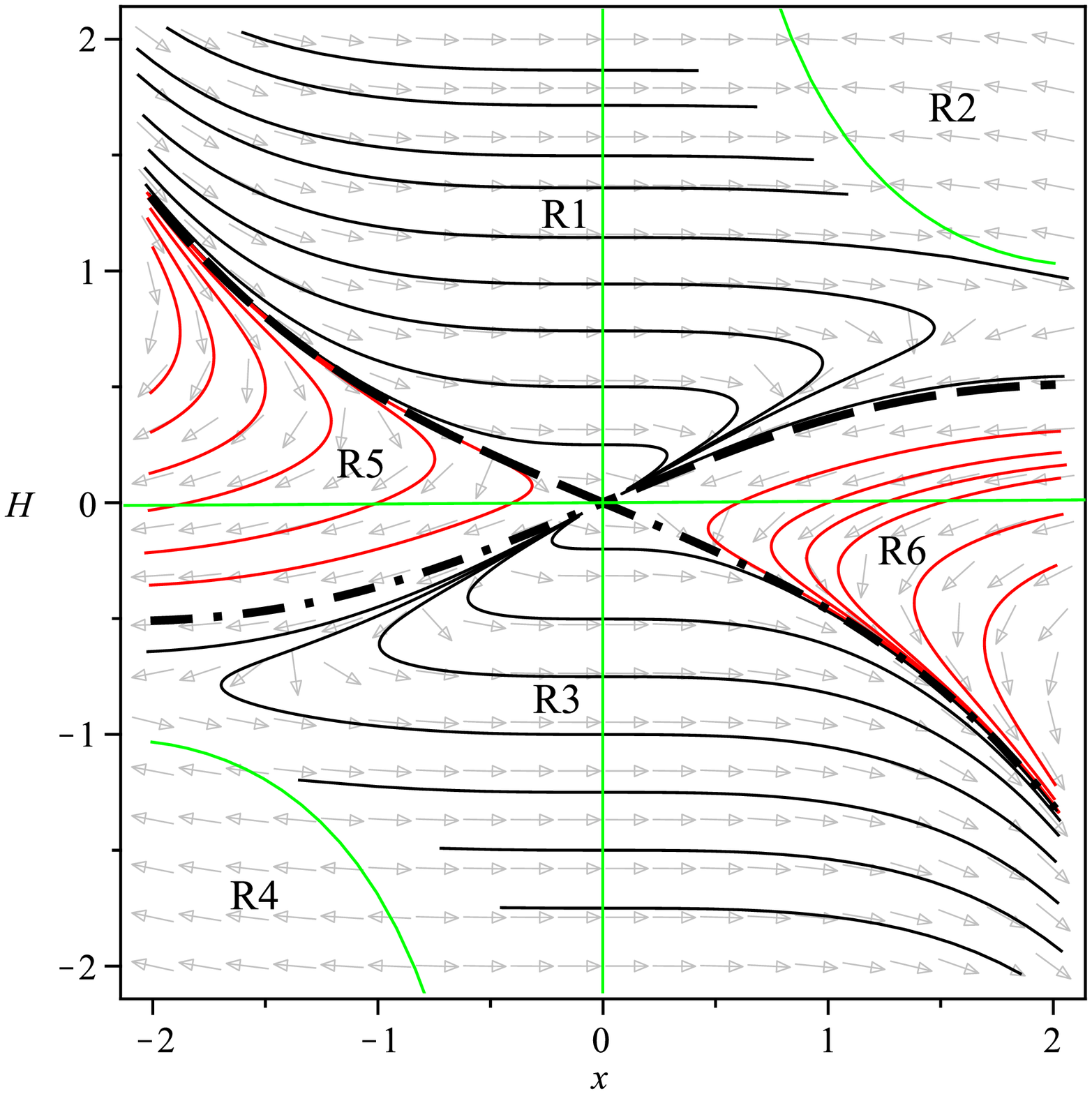}
\includegraphics[width=5.5cm]{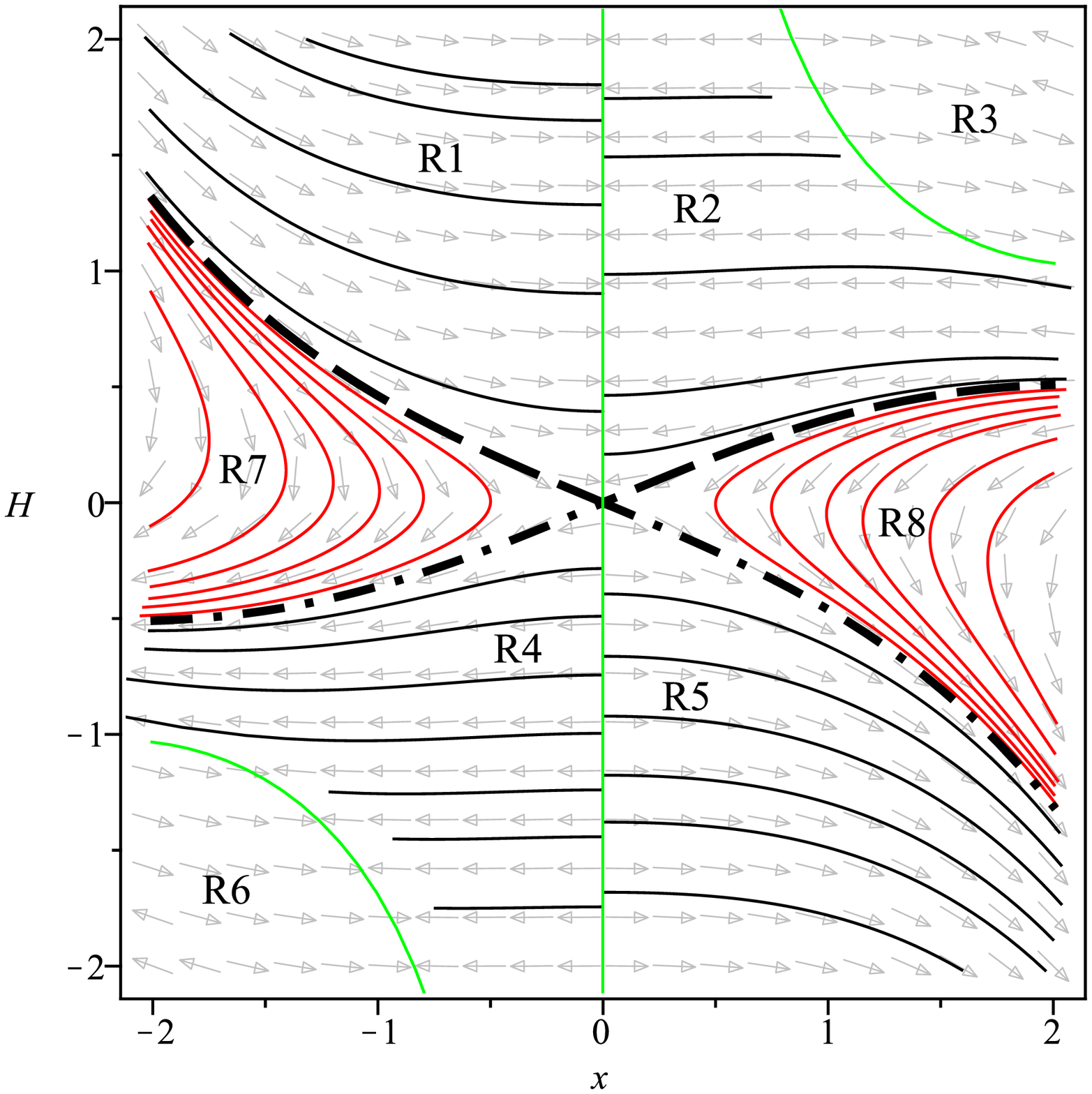}
\includegraphics[width=5.5cm]{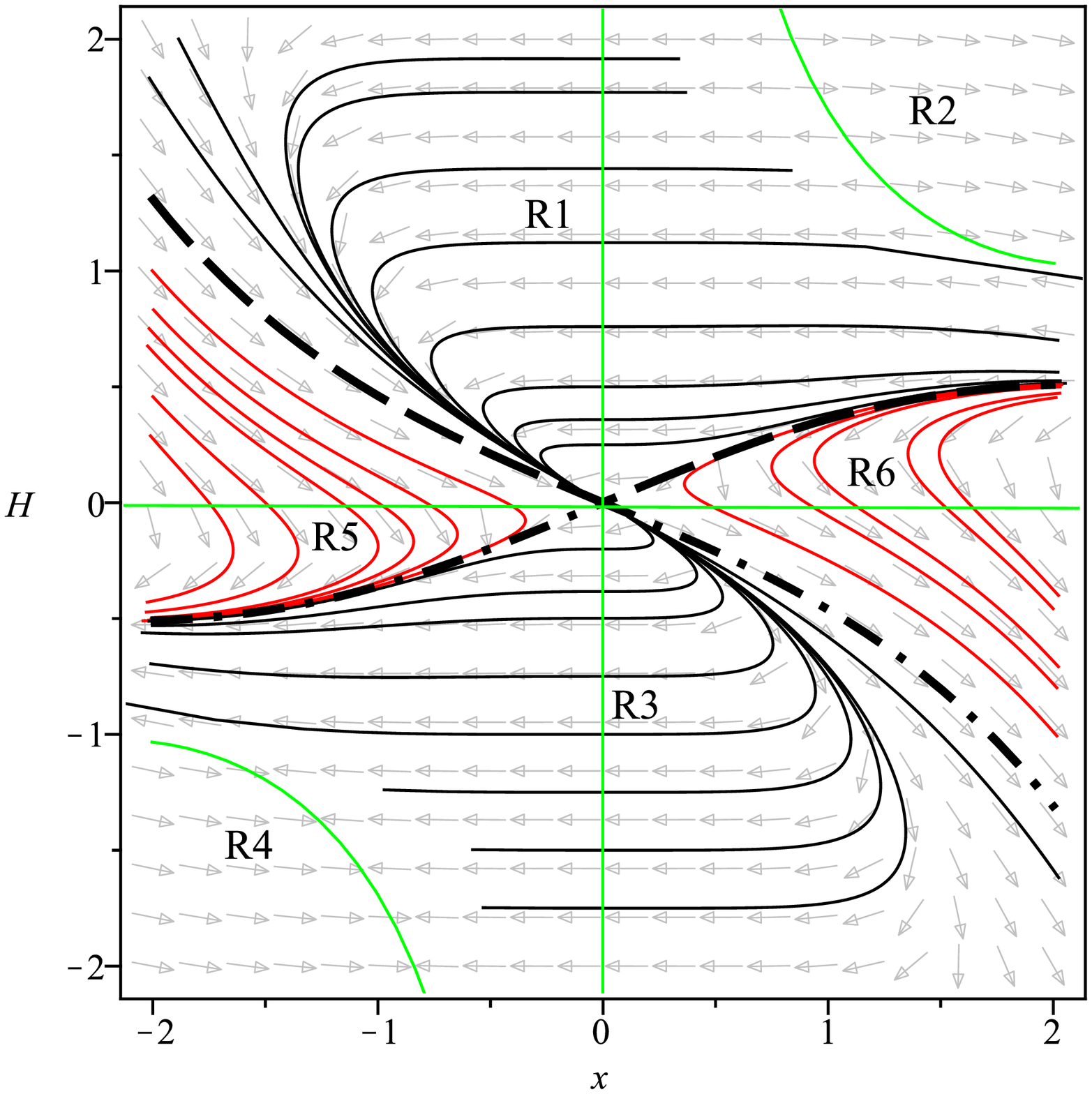}\vspace{0.7cm}
\caption{Phase portrait of the dynamical system \eqref{2d-dyn-syst} in the variables $X\equiv\dot\phi$ -- horizontal direction -- and $H$ (Hubble rate) -- vertical direction -- for the exponential potential $V=V_0\exp{(-\lambda\phi)}$, for different values of the parameter $\lambda$: From left to the right $\lambda=2$, $\lambda=0$ ($V=V_0$) and $\lambda=-2$, respectively. The thick dashed and thick dot-dashed curves represent the condition $V=0$ and are separatrices in the phase plane. The thin solid curves in the top right and in the bottom left corners are also separatrices. An additional separatrix arises in the middle figure: the vertical line $X=0$. The mentioned separatrices split the phase plane into non-connected regions R1, R2, R3, ..., etc. The regions R5 and R6 in the left and in the right -- regions R7 and R8 in the middle figure -- represent cosmological evolution driven by negative potential $V<0$.}\label{fig-2}\end{figure}

\subsubsection{The exponential potential}

As an illustration we consider the exponential potential that is a very useful self-interaction potential for applications in cosmology. In this case we have that $$V(\phi)=V_0\exp{(-\lambda\phi)},\;\;\lambda=\text{const.}$$ Hence, the dynamical system \eqref{dyn-syst} reduces to a two-dimensional autonomous system of ordinary differential equations (first and second equations in \eqref{dyn-syst}):

\bea &&\dot X=\frac{-3X(2H+\sigma_0 X^3)(1-3\sigma_0 XH)+\lambda\left(6H^2+6\sigma_0 X^3H-X^2\right)}{2-12\sigma_0 XH+3\sigma_0^2X^4},\nonumber\\
&&\dot H=-\frac{X^2(1-3\sigma_0 XH)(1-9\sigma_0 XH)+\sigma_0 X^2\lambda\left(3H^2+3\sigma_0 X^3H-X^2/2\right)}{2-12\sigma_0 XH+3\sigma_0^2X^4}.\label{2d-dyn-syst}\eea The corresponding phase space is the plane $(X,H)$. Although the phase plane is unbounded, here we concentrate in the region around the origin $(0,0)$, so that we do not need to draw the entire phase portrait.

In the figure FIG. \ref{fig-2} the phase portrait of the dynamical system \eqref{2d-dyn-syst} is shown for different values of the constant parameter $\lambda$. It is appreciated that there are several separatrices. The thick dash and dash-dot curves correspond to vanishing of the potential $V=0$ (see the right-hand equation in \eqref{cond-1}):

\bea H_\pm=-\frac{\sigma_0 X^3}{2}\pm\sqrt{\frac{\sigma_0^2 X^6}{4}+\frac{X^2}{6}}.\nonumber\eea Hence, the region above of the thick dashed curve, corresponding to $H>H_+$: i. e., the union of R1 and R2 in the left-hand and in the right-hand figures, respectively, and the union of regions R1, R2 and R3 in the middle figure, is for non-negative potentials $V\geq 0$ (the cases of interest in our paper). The same holds true for the region below of the thick dash-dotted curve, that corresponds to the condition $H<H_-$: i. e., the union of R3 and R4 in the left-hand and in the right-hand figures, respectively, and the union of regions R4, R5 and R6 in the middle figure. These correspond to fulfillment of the condition $V\geq 0$ as well. Meanwhile, the regions R5 and R6 in the left-hand and in the right-hand figures in FIG. \ref{fig-2}, and R7 and R8 in the middle figure, correspond to the cubic galileon with the negative potential $V<0$. 

The separatrices depicted by the thin solid curves in the top-righ and bottom-left corners of the figures in FIG. \ref{fig-2} are given by the condition:

\bea H=\frac{2+3\sigma_0^2 X^4}{12\sigma_0 X},\label{cond-3}\eea and correspond to vanishing of the denominator in the RHS of equations \eqref{dyn-syst} and, consequently, of \eqref{2d-dyn-syst}. This means that these curves represent asymptotic states where a sudden change in the orientation of the phase plane orbits happens. For the constant potential (middle figure in FIG. \ref{fig-2}) an additional separatrix arises: the vertical line $(0,H)$, that separates the region in the phase plane where $\dot\phi<0$ (left half), from the region where $\dot\phi>0$ (right half).

As seen from FIG. \ref{fig-2} all of the possible orbits that can cross the value $H=0$ -- including the possibility of a non-singular bounce whenever $\dot H>0$ -- lay in the regions R5, R6 (left and right) or R7, R8 (middle), i. e., these represent unphysical cosmological evolution driven by negative potentials $V<0$. Notice that, no matter whether the slope of the exponential potential is negative (for instance $\lambda=2$ in the left-hand figure in FIG. \ref{fig-2}) or positive ($\lambda=-2$ in the right-hand figure), orbits originated by the initial condition: $H(t_0)>H_+$, are attracted towards the late-time attractor at the origin: $(X,H)=(0,0)$. For orbits in the region with $H<H_-$, the origin is a past attractor instead, i. e., it is the starting point of the cosmic history.

\begin{figure}\centering
\includegraphics[width=5.5cm]{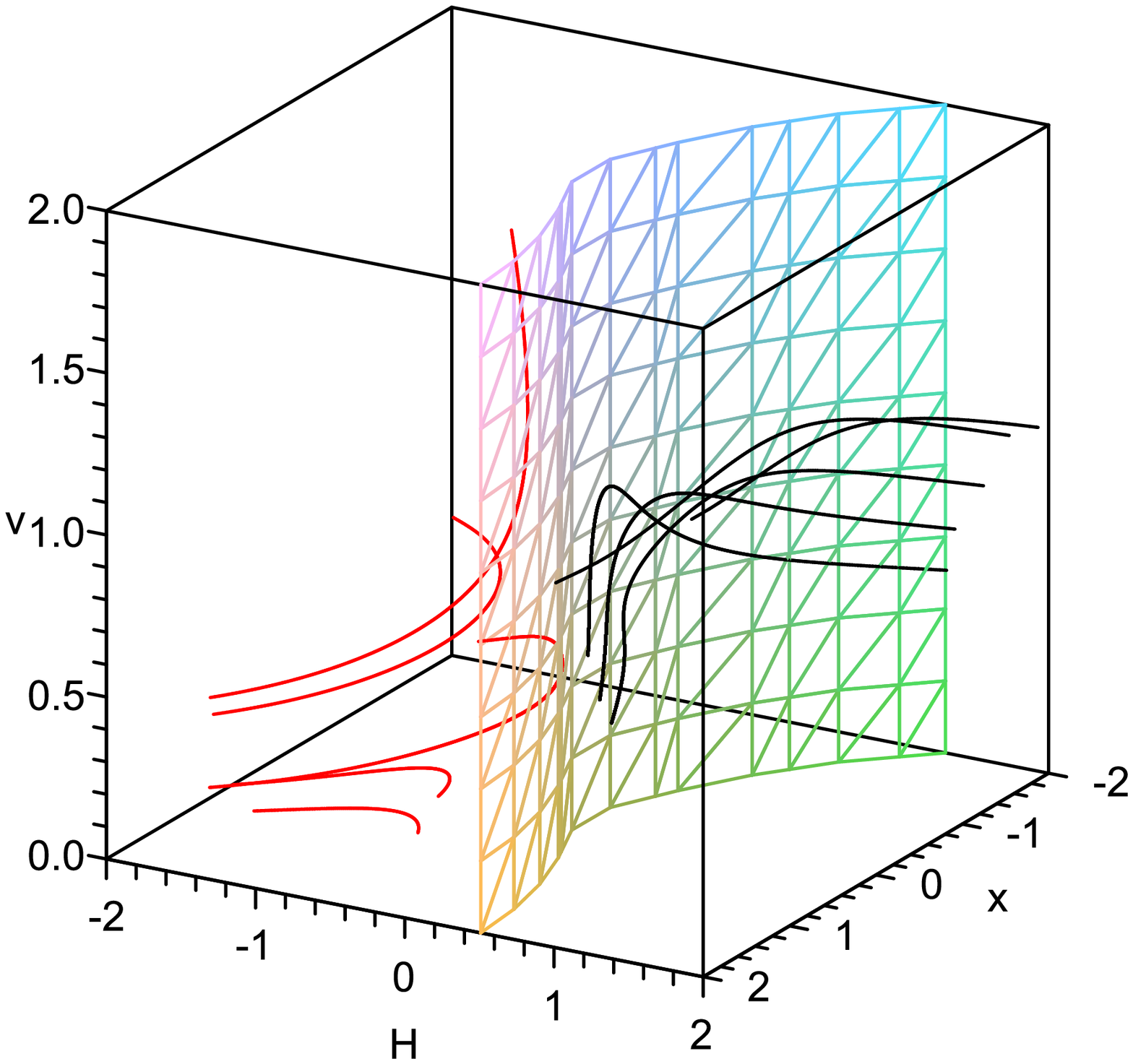}
\includegraphics[width=5.5cm]{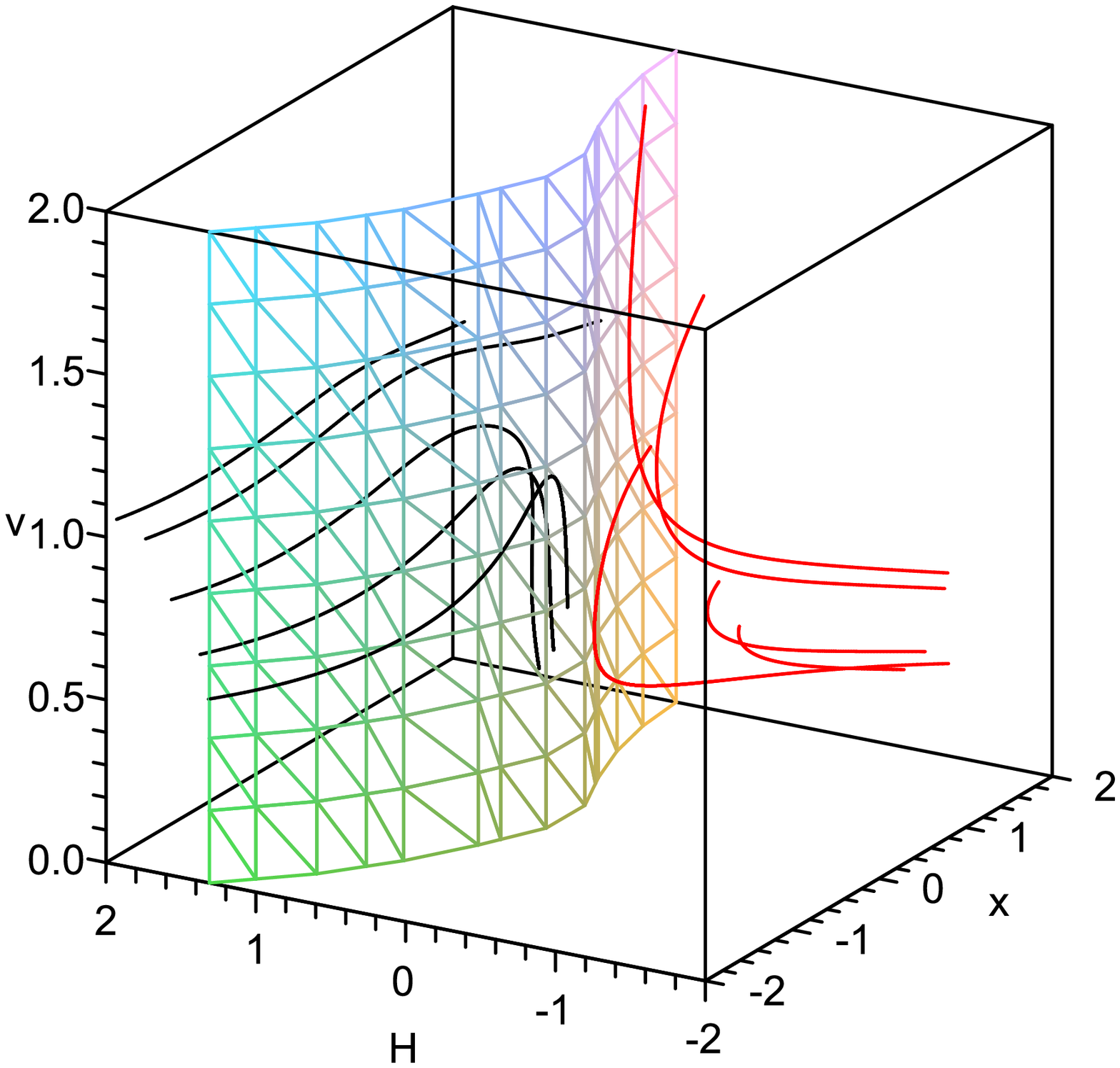}
\includegraphics[width=5.5cm]{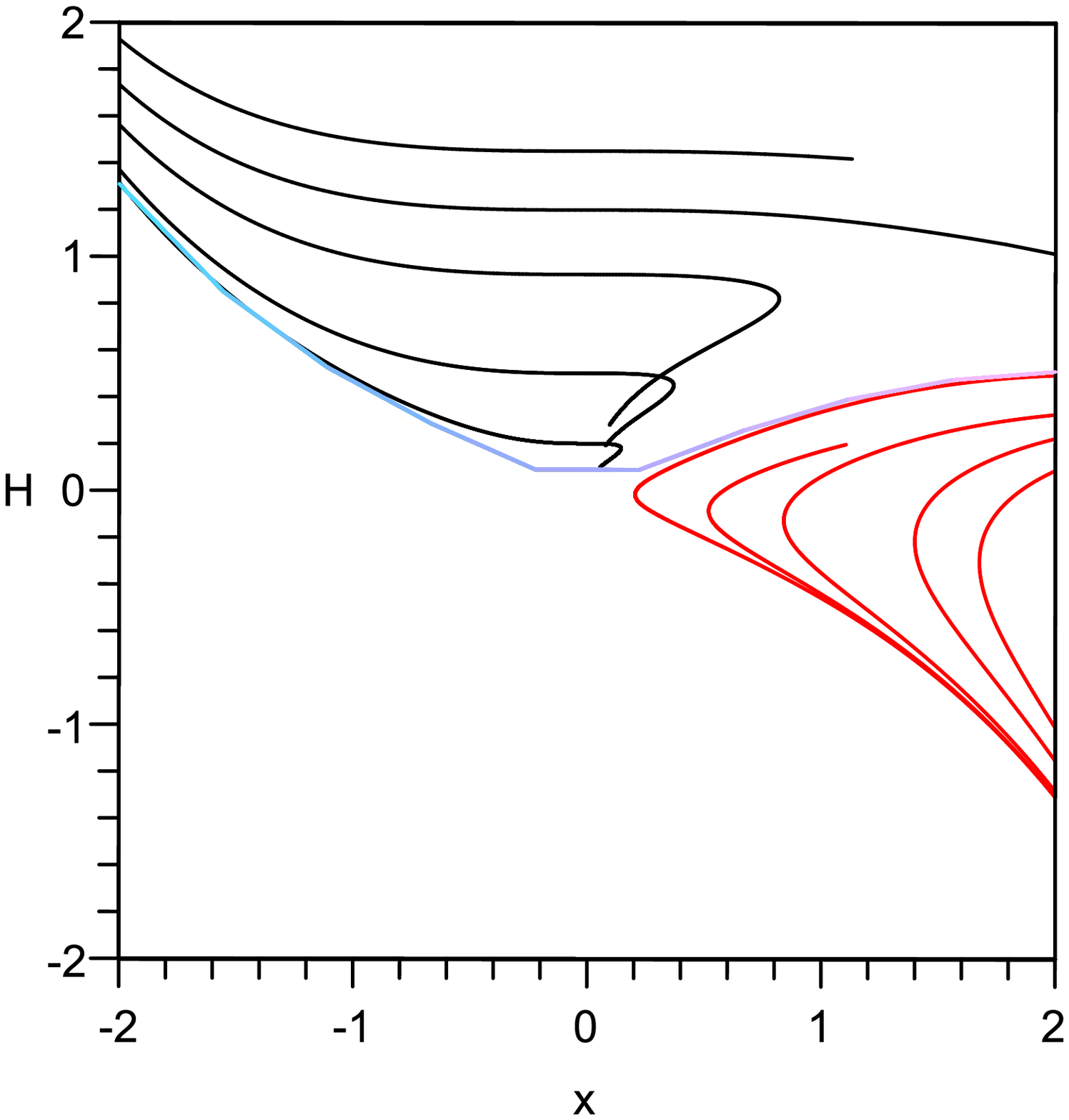}\vspace{0.7cm}
\caption{Phase portrait of the 3d dynamical system \eqref{dyn-syst} for the power-law potential $V=V_0\phi^{-p}$ with $p=2$, seen from different angles: $[\theta,\vphi]=[30^o,70^o]$ -- left, $[-150^o,70^o]$ -- middle, and the projection onto the plane $X,H$ -- right, respectively. We use the coordinates $X=\dot\phi$ and $H$ -- Hubble rate, while the additional coordinate $v$ -- vertical direction, represents the slope of the potential (it is properly the coordinate $\lambda$ in the main text). The solid dark curves represent orbits originated from initial data within the region where the potential is non-negative and $H>H_+$. These are either attracted by the lineal critical manifold ${\cal M}:(X,H,v)=(0,0,v)$, or go somewhere else, evolving completely within the region that is bounded by the condition $V\geq 0$, that is represented in the figure by the vertical surface separating the dark-colored orbits from the red ones. The orbits represented by the solid (red) curves do the cross of $H=0$, however, these arise only for negative potentials that are not of physical interest.}\label{fig-3}\end{figure}

\subsubsection{Power law potential}

For the power law potential \eqref{pwl-pot}: $V(\phi)=V_0\,\phi^{-p}$, the variable $\lambda$ is a function of $\phi$. In this case the phase space is a three-dimensional manifold $(X,H,\lambda)\in\text{R}^3$. The dynamical system corresponding to this case is properly \eqref{dyn-syst} with $f(\lambda)=\lambda^2/p$. In the figure FIG. \ref{fig-3} several orbits of \eqref{dyn-syst} are drawn for $p=2$. Those originated from initial conditions in $H>H_+$ ($V>0$) -- dark curves -- do not cross the value $H=0$, while those in the region where $V<0$ -- red curves -- do the crossing. However, here we consider physical situations where $V\geq 0$, so that the red orbits represent cosmological behavior without physical interest. In the figures the vertical surface that separates the orbits driven by $V>0$ from orbits driven by $V<0$, corresponds to the condition $V=0$ and, in implicit form, is given by: $$H+\frac{\sigma_0 X^3}{2}-\sqrt{\frac{\sigma_0^2 X^6}{4}+\frac{X^2}{6}}=0.$$

%%%%%%%%%%%%%%%%%%%%%%%%%%%%%%%%%%%%%%%

\section{Conclusion}\label{sec-conclu}

In the present paper we have generalized the results of \cite{rtgui} for potentials beyond the exponential one. Here we have considered four other potentials which are of cosmological interest. The conclusion that the vacuum dynamics of the cubic galileon is not fully contained in the -- apparently more general -- dynamics of the cubic galileon with the background matter, seems to be independent of the specific functional form of the self-interaction potential. As we have shown, this happens to be a phase space manifestation of a cosmological version of the vDVZ discontinuity, not previously found in the bibliography. Even in the paper \cite{rtgui}, where a similar result was found for the exponential potential case, the effect was not related with the above mentioned discontinuity. The natural resolution of the vDVZ discontinuity is given by the cosmological Vainshtein screening mechanism that is typical of theories with the cubic self-interaction $\nabla^2\phi(\der\phi)^2$.

\subsection{Generality of the results}

Although in this paper we have considered specific self-interaction potentials for the cubic galileon, our main finding: that there are late-time (attractor) solutions in the vacuum case which are not found in the presence of matter degrees of freedom other than the galileon itself, is a quite general result. Actually, as we have shown in section \ref{sec5}, the vacuum equilibrium state represented by the critical point $P_{6v}$ -- being a local future attractor -- exists for monotonically growing potentials independent on their specific functional form $V=V(\phi)$. An example can be the growing exponential or the power-law with positive power $\propto\phi^{2k}$ ($k>0$). This phantom-like attractor can appreciably modify the late-time cosmic dynamics by placing a big rip event at the end of the cosmic history. 

Other vacuum solutions that are also erased through the cosmological Vainshtein screening mechanism: the critical points $P_{7v}$ and $P_{8v}$, exist for vanishing potentials or for potentials that vanish asymptotically such as, for instance, the decaying exponential and the inverse power-law (assuming that $\dot\phi>0$, i. e., that the galileon is a monotonically increasing function of the cosmic time). These are associated with contracting solutions and are local sources (past attractors), so that they do not modify the late-time cosmic dynamics in any appreciable way.

The above results entail that, for the potentials of interest for the quintessence models, i. e., those that decay with the cosmic expansion and yield to the expected late-time dynamics, the phantom behavior does not arise, meaning that in general the late-time cosmic dynamics is not modified by the cubic self-interaction. In this sense the vDVZ discontinuity and the occurrence of the Vainshtein cosmic screening are effects that can be met only at early times (solutions $P_{7v}$ and $P_{8v}$) where the cubic self-interaction may become appreciable. This, in turn, may be an indication that the evolution of our universe might nest a previous stage of contracting evolution that evolved into the expanding phase through the static universe (no bounce). We shall give further arguments on this issue in a forthcoming publication.

%%%%%%%%%%%%%%%%%%%%%%%%%%%%%%%%%%%%%%%%%

\section*{Acknowledgments}

RDA, TG, UN and IQ want to thank SNI-CONACyT and IAC of M\'exico for continuous support of their research activity, meanwhile GL thanks to Department of Mathematics and to Vicerrector\'ia de Investigaci\'on y Desarrollo Tecnol\'ogico at Universidad Catolica del Norte for financial support. RDA and UN also acknowledge PRODEP and CIC-UMSNH.

%%%%%%%%%%%%%%%%%%%%%%%%%%%%%%%%%%%%%%%%%%%%%%%%%%%%%%%%%%%%%%%%%%%%%%%%%%%%%%%%%%%%%%%%%%%%%%%%%%%%%%%%%%%%%%%%%%%%%%%%%%%%%%%%%%%%%%%%%

\end{document}